\documentclass[10pt,a4paper]{scrartcl}
\usepackage[left=13mm,right=13mm, top=17mm, bottom=27mm]{geometry}
\usepackage[utf8]{inputenc}
\usepackage[T1]{fontenc}
\usepackage[english]{babel}
\usepackage{csquotes}
\usepackage{amsmath}
\usepackage{amssymb}
\usepackage{graphicx}
\usepackage{epsfig}
\usepackage{epstopdf}
\usepackage{amsthm}
\usepackage{enumerate}
\usepackage{color}
\usepackage{url}
\usepackage{here}
\usepackage{caption}
\usepackage{subcaption}
\usepackage{rotating}
\usepackage{tabularx}
\usepackage{appendix}
\usepackage{mathtools}
\usepackage{tabu}
\usepackage[capitalize,nameinlink,noabbrev]{cleveref}
\usepackage{autonum}
\addtokomafont{sectioning}{\rmfamily}
\usepackage[colorinlistoftodos, backgroundcolor=yellow, linecolor=yellow]{todonotes}

\newtheorem{theorem}{Theorem}[section]
\newtheorem{definition}[theorem]{Definition}

\newtheorem{lemma}[theorem]{Lemma}

\newcommand{\R}{\mathbb{R}}

\newcommand{\B}{\mathbb{B}}

\newcommand{\ph}{\varphi}

\DeclareMathOperator{\erf}{erf}

\newcommand{\bb}{\mathcal{B}}

\newcommand{\dx}{\mathrm{d}x}
\newcommand{\dy}{\mathrm{d}y}

\newcommand{\Si}{\mathrm{Si}}

\newcommand{\St}{\mathrm{St}}
\newcommand{\CN}{\mathrm{CN}}

\newcommand{\reg}{\mathrm{reg}}
\newcommand{\pe}{\mathrm{pe}}
\newcommand{\new}{\mathrm{new}}

\newcommand{\x}{\|x\|}
\newcommand{\y}{\|y\|}
\newcommand{\Ei}{\mathrm{Ei}}
\begin{document}
\setlength{\parindent}{0pt}
\begin{center}
\textbf{\Large Decorrelation of poroelastic data via multiscale mollifiers wavelets}
\end{center}
\begin{center}
B. Kretz\footnote[1]{Geomathematics Group, Department of Mathematics, University of Siegen, Emmy-Noether-Campus,
	Walter-Flex-Str. 3, 57068 Siegen, Germany, E-Mail addresses: kretz@mathematik.uni-siegen.de, michel@mathematik.uni-siegen.de},
	W. Freeden\footnote[2]{Geomathematics Group, University of Kaiserslautern, Paul-Ehrlich-Straße,
		67663 Kaiserslautern, Germany, E-Mail address: freeden@mathematik.uni-kl.de},
V. Michel\footnotemark[1]
\end{center}

\begin{center}
\textbf{Abstract}
\end{center}
\begin{abstract}
Poroelasticity can be classified with geophysics and describes the interaction between solids deformation and the pore pressure in a porous medium.
The investigation of this effect is anywhere interesting where a porous medium and a fluid come together into play, for example this is the case in geothermics. More precisely, it is an important aspect in reservoir management since the replacement of the water in the reservoir some kilometers below the Earth's surface has an effect on the surrounding material and of course displacement of the solid increases or decreases the pore pressure.\\
The underlying physical processes are deduced with the help of linear elasticity, conservation of linear momentum, conservation of mass and Darcy's law.
They result in partial differential equations, called the quasistatic equations of poroelasticity (QEP).\\
In this paper, we want to do a multiscale decomposition of the components displacement and pore pressure. This should provide us with more information about the data that means visualize underlying structures in the different decomposition scales that cannot be seen in the whole data. The aim is to detect interfaces and extract more details of the data.\\
For this purpose, we construct physically motivated scaling functions by mollifying the appropriate fundamental solutions. Here we have a closer look at the scaling functions fulfilling the necessary theoretical requirements of an approximate identity. The corresponding wavelets are constructed by subtraction of two consecutive scaling functions.
\end{abstract}



\section{Introduction}
Poroelasticity is an extension of the classical theory of linear elasticity and can be dated back to Biot in the 1930s (see \cite{biot1935}). It is everywhere interesting where a porous medium and a fluid come into play because it describes the effect of solids deformation on pore pressure and vice versa. In geothermics for example, these processes are considerable since aquifers are assumed to be poroelastic (see for example \cite{ghasssemi_tao, wang} for the treatment of poroelasticity in reservoirs). The extraction of water out of an aquifer and the reinjection afterwards have an influence on the surrounding material right up to formation of fissures which can cause for example seismic events. Also solid deformation surely brings variety to the pore pressure. These influences can be interesting during drilling and in addition during the exploitation phase. 

Our aim is to construct scaling functions and wavelets that have a physical relevance to the underlying partial differential equations due to their fundamental solutions. This is done by applying a so-called mollifier technique to the fundamental solutions. Both, mollifiers and wavelets have their origin long time ago and are used in many research topics.

In general, mollifiers are smooth auxiliary functions whose beginnings go back to Friedrichs in 1944 (see \cite{friedrichs}). The field for the usage of mollifiers is very broad and we can only point out some applications to show how widespread their application is. Under the aspect of inverse problems, mollifiers can be found in \cite{engl_hanke_neubauer, engl_louis_rundell, louis, louismaass} with the focus on operator solutions. Furthermore there are applications of mollifiers with finite elements in \cite{febriantoetal}, for spherical deconvolutions in \cite{hielscher_quellmalz}, for optimization problems in \cite{jongen_stein}, for the Laplace transform in \cite{schuster2002}, for vector tomography in \cite{schuster2001}, for the solution of nonlinear systems of differential equations in \cite{wang_oberguggenberger} and for the recovery of piecewise analytical functions in \cite{tanner}.

The theory of wavelets began earlier (see \cite{haar}) since the Haar wavelets can be seen as the most simple wavelets. Several ways to construct wavelets should be summarized here in short. All approaches have as a common feature the construction of wavelets as basis functions for multiscale analysis. For main topics of wavelets we refer for example to \cite{chui, daubechies, louismaassrieder, mallat2009}. Together with the Haar wavelet, we can mention the Daubechies wavelet (see \cite{daubechies}) and the Meyer wavelet (see \cite{meyer}) as wavelets on the real line. For the construction of wavelets on the sphere there exist many basic approaches: a tensor product ansatz (\cite{dahlkemaass}), a group-theoretic approach (\cite{antoineetal2002}), kernel wavelets (\cite{holschneider1996, lain2003, narcowichward, pottstasche, sweldens, wiauxetal}), frequency reflected wavelets (\cite{freedenschreiner1998, freeden_windheuser}), layer potentials (\cite{freeden_gerhards2010, freeden_mayer}), boundary integral equations (\cite{freedenmayer2006, harbechtschneider, kunothsahner}) and approaches with focus on localization (\cite{lain2007, michel2011}). For a more detailed overview about wavelets see \cite{freeden2020, michel}.

The idea of the approach which we pursue here in poroelasticity goes back to \cite{freeden_schreiner2006} and is also performed for example in \cite{blick2015, blick_eberle, freeden2021, freeden_blick}. We want to construct mollifier wavelets which have a connection to the underlying physical problem and should be able to decompose geophysical quantities. Mollifier wavelets like that were constructed for the gravitational potential problem on the one hand to reconstruct the potential from density data and on the other hand for the decomposition of gravity or density data in different bands. These approaches have the aim to detect underlying structures and interfaces that cannot be seen in the whole data. Our focus here is on the decomposition of the relevant poroelastic quantities displacement and pore pressure. The approach is as follows: We start with the related partial differential equations. The important key element are the fundamental solutions associated to the partial differential equations. They have singularities which we want to avoid by mollifying them with a Taylor expansion. The degree of the mollification can be controlled by a scaling parameter which corresponds later to the different bands of the decorrelation. Applying now the poroelastic differential operator on these mollified fundamental solutions, results in the so-called source scaling functions. These are the functions which we are interested in for performing the decomposition. They fulfill the theoretic property of an approximate identity, which is the basis and necessary for all further steps. The wavelets are obtained as the difference between two consecutive scaling functions. The wavelets can be understood as band-pass filters whereas the scaling functions as low-pass filters. 

\section{Poroelasticity: equations and fundamental solutions}
The aim of this section is to derive the quasistatic equations of poroelasticity (QEP) and their fundamental solutions very shortly. Beforehand we start with some important basics.
$\B_r^n(x)$\index{ ball@$\B_r^n(x)$} is defined as the closed ball with center $x \in \R^n$ and radius $r>0$ that means
\begin{align}
	\B_r^n(x) \coloneqq \{ y \in \R^n \big| \|x-y\| \leq r \}.
\end{align}
We denote by $\bb$ a regular region, that means $\bb$ is a subset of $\R^n$ which is open and connected and additionally $\bb$ is bounded and $\partial \bb$ is an orientable piecewise smooth Lipschitzian manifold of dimension $n-1$ (see \cite{freeden_latticepoint}).\\
For a tensor-valued function $\boldsymbol{f}$, the divergence is meant row-wise, that means it is applied in the following way
\begin{align}
	 \nabla_x \cdot \boldsymbol{f}= \left( \sum_{j=1}^n \frac{\partial}{\partial x_j} f_{ij} \right)_{i=1,\dots,n},\label{def:divergence}
\end{align}
where $f_{ij}$ is the component of $\boldsymbol{f}$ in row $i$ and column $j$.
Furthermore, we need some special functions, namely the error function and the exponential integral. 
\begin{definition} \label{def:erf}
	The error function (also called Gauss error function) is given by
	\begin{align}
		\erf (x) \coloneqq \frac{2}{\sqrt{\pi}} \int_0^x e^{-\tau^2} \ \mathrm{d} \tau. \label{eq:errorfunction}
	\end{align}	
	In this work, we use the error function for real arguments.
\end{definition}
\begin{definition} \label{def:Ei}
	The exponential integral has to be understood in the sense of the Cauchy principal value and is defined as the following integral
	\begin{align}
		\mathrm{Ei}(x)\coloneqq -\int_{-x}^\infty \frac{e^{-t}}{t} \ \mathrm{d}t=\int_{-\infty}^x \frac{e^t}{t} \ \mathrm{d}t, \quad x>0.
	\end{align}
Furthermore, the following characterization is given as an alternative representation
\begin{align}
	\mathrm{Ei} (x)=C + \ln |x| +\sum_{k=1}^\infty \frac{x^k}{k \cdot k!}, \quad x\in \R\setminus \left\{ 0 \right\},
\end{align}
with $C$ as the Euler constant.
\end{definition}
Hence we have the property
\begin{align}
	\mathrm{Ei}(-t) \cdot t^{k} \to 0 \text{ for } t \to 0 \text{ for } k\geq 1,	\label{eq:Eilimes}
\end{align}
since $x \cdot \ln |x| \to 0\ (x \to 0)$,
which we need later for some limit considerations.

Now we come to some integrals in connection with the exponential integral that we need for the proof of the necessary theoretical properties of our desired scaling functions.
\begin{lemma}\label{remark:exp_integral}
	The following integrals are useful for the integration of the time-dependent scaling functions later
	(see \cite{gradshtein_ryshik}):
	
	\begin{align}
		&\int \exp \left(-\frac{\tau^2}{4 C_2 t}  \right) \ \mathrm{d} t \phantom{\cdot\ \frac{1}{t}}= \frac{\tau^2 \mathrm{Ei}\left(-\frac{\tau^2}{4 C_2 t}  \right)}{4 C_2}+t\cdot \exp \left(-\frac{\tau^2}{4 C_2 t}  \right)\\
		&\int \exp \left(-\frac{\tau^2}{4 C_2 t}  \right)\cdot \frac{1}{t}\ \mathrm{d} t = -\mathrm{Ei}\left(-\frac{\tau^2}{4 C_2 t}  \right)\\
		&\int \exp \left(-\frac{\tau^2}{4 C_2 t}  \right)\cdot \frac{1}{t^2}\ \mathrm{d} t = \frac{4 C_2  \exp \left(-\frac{\tau^2}{4 C_2 t}  \right)}{\tau^2} \\
		&\int \exp \left(-\frac{\tau^2}{4 C_2 t}  \right)\cdot \frac{1}{t^3}\ \mathrm{d} t = \frac{4 C_2  \exp \left(-\frac{\tau^2}{4 C_2 t}  \right) (4 C_2 t +\tau^2)}{t \tau^4}\\
		&\int \exp \left(-\frac{\tau^2}{4 C_2 t}  \right)\cdot \frac{1}{t^4}\ \mathrm{d} t = \frac{4 C_2  \exp \left(-\frac{\tau^2}{4 C_2 t}  \right) (32 C_2^2t^2+8 C_2t \tau^2+\tau^4)}{t^2 \tau^6}.
	\end{align}
\end{lemma}
Our last point, to mention for the basics, is the convolution between a function containing a delta distribution and a data function $f$.
With the general properties of distributions (see for example \cite{hoermander, jantscher, kythe}), especially in combination with derivatives and convolutions, we have a look at distributions of the form $\phi(x)\delta_t$ or $\phi(x)\delta'_t$ with a distributional time dependent part (in our case the delta distribution). In these cases, we set $\delta_t f(y,\cdot) \coloneqq f(y,t)$ and accordingly define the convolution of these (scaling) functions with the data function $f$ (assumed to be differentiable with respect to $t$) as
\begin{align}
	\left( \phi \delta_t \ast f \right) &\coloneqq \int_{\R^2} \phi(x-y) \left( \delta_t f(y,\cdot) \right)\ \dy = \int_{\R^2} \phi(x-y) f(y,t)\ \mathrm{d}y, \label{eq:conv_deltat}\\
	\left(	\phi \delta'_t \ast f \right) &\coloneqq \left(\phi \delta_t \ast \frac{\partial}{\partial t}f \right)= \int_{\R^2} \phi(x-y) \frac{\partial}{\partial t}f(y,t)\ \mathrm{d}y. \label{eq:conv_deltastricht}
\end{align}

The quasistatic equations of poroelasticity (QEP) are well-known (see \cite{cheng, cheng_detournay})		
Their derivation is based on the physical laws stress-strain relation, equilibrium equation, fluid mass balance equation and Darcy's law.

The QEP in dimensionless form are given by
\begin{align}
	-\frac{\lambda+\mu}{\mu} \nabla_x (\nabla_x \cdot u)-\nabla_x^2 u+\alpha \nabla_x p &= 0, \label{eq:qep11}\\
	\partial_t (c_0 \mu p+\alpha(\nabla_x \cdot u))-\nabla_x^2 p &= 0. 
\end{align}

For them, a fundamental solution tensor $\boldsymbol{G}$ in $u$ and $p$ is given (see \cite{augustin2012, augustin2015}) by
\begin{align}
	\boldsymbol{G} (x,t)= \begin{pmatrix}
		\boldsymbol{u}^{\CN}(x) \delta_t & p^{\St}(x)\delta_t  \\
		u^{\Si} (x,t)& p^{\Si}(x,t)
	\end{pmatrix}
\end{align}
with the following entries\index{fundamental solution!poroelasticity!homogeneous}
\begin{align}
	p^{\Si}(x,t)&=\frac{1}{4 \pi t} \exp \left( -\frac{\|x\|^2}{4 C_2 t} \right), \label{eq:pSi}\\
	u^{\Si}(x,t)&= C_1 \frac{x}{2 \pi \|x\|^2} \left( 1-\exp \left( -\frac{\|x\|^2}{4 C_2 t} \right) \right), \\
	p^{\St}(x)&= C_1 \frac{x}{2 \pi \|x\|^2},\\
	\boldsymbol{u}_{ki}^{\CN}(x)&= C_3 \frac{1}{2 \pi} \left( -\delta_{ki}\ln (\|x\|)+C_4 \frac{x_i x_k}{\|x\|^2} \right)
\end{align}
and the abbreviated constants
\begin{align}
	C_1\coloneqq &\frac{\alpha}{c_0 (\lambda+2\mu)+\alpha^2},\ && \hspace*{-2cm} C_2\coloneqq \frac{\lambda+2\mu}{c_0 \mu (\lambda+2\mu)+\mu \alpha^2}, \label{eq:C1C2} \\
	C_3\coloneqq &\frac{c_0(\lambda+3 \mu)+\alpha^2}{2(c_0(\lambda+2 \mu)+\alpha^2)}, \ && \hspace*{-2cm} C_4\coloneqq \frac{c_0(\lambda+\mu)+\alpha^2}{c_0(\lambda+3\mu)+\alpha^2}. \label{eq:C3C4}
\end{align}

\section{Mollification of the fundamental solutions}
Before we go over to the fundamental solutions in poroelasticity, we want to show the mollification scheme that we use for a general function.
We assume a function $F$ which can be represented in the following way $F(x)=f(\|x\|)g(x_1)h(x_2)$ and we only regularize the radially symmetric function $f$. This is done in the following way: We do a Taylor expansion in $u_0$ for $f(\sqrt{u})$ for the case $\|x\|<\tau$ which is given by
\begin{align}
	f_\reg(u) \coloneqq f(\sqrt{u_0})+\left( \frac{\partial}{\partial u} f(\sqrt{u}) \right) \bigg|_{u=u_0}(u-u_0).
\end{align}
Now substituting $u=\|x\|^2$ and $u_0=\tau^2$ yields
\begin{align}	
	f_\reg(x)=f(\tau)+f'(\tau)\frac{1}{2\tau}(\|x\|^2-\tau^2).
\end{align}
We call the composite function $F_\tau$ the Taylor mollification of $F$, that means
\begin{align}
	F_\tau(x)\coloneqq \begin{cases}
		f(\|x\|)g(x_1)h(x_2),\quad & \|x\| >\tau,\\
		(f(\tau)+f'(\tau) \frac{1}{2\tau}(\|x\|^2-\tau^2))g(x_1)h(x_2),\quad & \|x\| <\tau.
	\end{cases} \label{eq:F_tau}
\end{align}
It is easy to verify the following theorem.
\begin{theorem} \label{theorem:reg_general}
	For continuously differentiable functions $f$, $g$ and $h$, the regularization $F_\tau$ in \eqref{eq:F_tau} is also continuously differentiable.
\end{theorem}

Now, we want to mollify each fundamental solution separately. We start with $p^\St$ and write it as a gradient in the following way (this is the only one of our functions that can be written in such a way)
\begin{align}
	\frac{x}{\|x\|^2}=\frac{1}{2} \nabla_x \left( \ln \left( \|x\|^2 \right) \right).
\end{align}
With this representation we have a radially symmetric factor which we want to mollify with the help of a Taylor expansion. Since we are interested in a regularization up to the linear term for $p^\St$, we have to do the Taylor expansion for the function above up to the quadratic term. By writing $\|x\|^2=u$, we obtain the derivative
\begin{align}
	\frac{1}{2} \ln u&=\frac{1}{2} \left( \ln u_0+\frac{1}{u_0}(u-u_0)-\frac{1}{2u_0^2}(u-u_0)^2 \right)+\mathcal{O}\left( (u-u_0)^3\right) \notag \\
	&= \frac{1}{2} \left( \ln u_0 +\frac{2u}{u_0}-\frac{3}{2}-\frac{u^2}{2u_0^2}  \right)+\mathcal{O}\left( (u-u_0)^3\right) \quad \text{as } u\to u_0.
\end{align}
Now setting $u_0$ as $\tau^2$ and $u$ as $r^2$ respectively $\|x\|^2$, we get the regularized fundamental solution
\begin{align}
	p^\St_{\reg}(x)\coloneqq \frac{C_1}{2\pi}\cdot \frac{1}{2} \nabla_x \left( \ln \tau^2+\frac{2 \|x\|^2}{\tau^2}-\frac{3}{2}-\frac{\|x\|^4}{2\tau^4} \right)
	= \frac{C_1}{2\pi}x \left( \frac{2}{\tau^2}-\frac{\|x\|^2}{\tau^4} \right).
\end{align}
We say that $\tau$ here is the regularization parameter.
We obtain the mollified fundamental solution with a distinction of cases in analogy to the general function above and get
\begin{align}
	p^{\St}_{\tau} (x) \coloneqq \begin{cases}
		\frac{C_1}{2\pi}\frac{x}{\|x\|^2},\ &\|x\| \geq \tau, \\
		\frac{C_1}{2\pi} \frac{x}{\tau^2} \left( 2- \frac{\|x\|^2}{\tau^2} \right),\ &\|x\| < \tau.
	\end{cases}
\end{align}

In preparation for the fundamental solution $u^\CN$ and for the proof of a weak convergence of the mollified fundamental solution to the fundamental solution, we consider the mollifications of the functions $L(x)=\frac{1}{\|x\|^2}$ and $N(x)=-\ln \|x\|$ given by
\begin{align}
	L_\tau(x)\coloneqq \begin{cases}
		\frac{1}{\|x\|^2},\quad & \|x\| \geq \tau,\\
		\left(\frac{2}{\tau^2}-\frac{\|x\|^2}{\tau^4}\right),\quad & \|x\| <\tau,
	\end{cases} \qquad
	N_\tau(x)\coloneqq \begin{cases}
		-\ln \|x\|,\quad & \|x\| \geq \tau,\\
		-\ln \tau -\frac{\|x\|^2}{2\tau^2}+\frac{1}{2},\quad & \|x\| <\tau.
	\end{cases}
\end{align}
 According to \cref{theorem:reg_general}, $L$ and $N$ are continuously differentiable. Moreover, they are non-negative as we will show now.
 \begin{lemma} \label{theorem:L-L_tau}
 	The mollifications of $L(x)=\frac{1}{\|x\|^2}$ and $N(x)=-\ln \|x\|$ given above, fulfill the property $(L-L_\tau)(x) \geq 0$ and $(N-N_\tau)(x) \geq 0$ for all $x\in\R^2 \setminus \{0\}$.
 	\begin{proof}
 		The term $(L-L_\tau)(x)=\frac{1}{\|x\|^2}-\frac{2}{\tau^2}+\frac{\|x\|^2}{\tau^4}$ is only relevant for the case $\|x\|  < \tau$ because for $\|x\| \geq \tau$ the difference is zero. The difference of the functions is radially symmetric and because of this reason, we differentiate the term with respect to $\|x\|=r$ and obtain
 		\begin{align}
 			\frac{\partial}{\partial r}  (L-L_\tau)(x)&=-\frac{2}{r^3}+\frac{2r}{\tau^4}.
 		\end{align}
 		With the following estimate
 		\begin{align}
 			r<\tau \ \Leftrightarrow \ 2r^4 < 2\tau^4 \ \Leftrightarrow \ \frac{2}{r^3} > \frac{2r}{\tau^4} \ \Leftrightarrow \ -\frac{2}{r^3}+\frac{2r}{\tau^4} <0,
 		\end{align}
 		we gain that $(L-L_\tau)(x)$ is a monotonically decreasing function with respect to the radial component $r=\|x\|$. With the fact that it has a zero value at $\|x\|=\tau$, we obtain $(L-L_\tau)(x) \geq 0$, which is a part of our desired result.
 		Differentiating the term $(N-N_\tau)(x)=-\ln\|x\|+\ln\tau+\frac{\|x\|^2}{2\tau^2}-\frac{1}{2}$ (also for $\|x\| < \tau$) for $\|x\|=r$ with respect to $r$, we get
 		\begin{align}
 			\ \frac{\partial}{\partial r} (N-N_\tau)(x)&=	-\frac{1}{r}+\frac{r}{\tau^2}.
 		\end{align}
 		Again an estimation gives us the following result
 		\begin{align}
 			r < \tau \ \Leftrightarrow \ \frac{r}{\tau^2} <\frac{\tau}{\tau^2}=\frac{1}{\tau} < \frac{1}{r} \
 			\Leftrightarrow \  -\frac{1}{r}+\frac{r}{\tau^2} <0. \notag
 		\end{align}
 		With the same argumentation as above, we obtain $(N-N_\tau)(x) \geq 0$.
 	\end{proof}
 \end{lemma}
Now we can prove the weak convergence in the following theorem.
\begin{theorem} \label{theorem:pSi-pSitau}
	Let us assume that $\mathcal{B}$ is a regular region in $\R^2$ and $f:\overline{\bb} \to \R^2$ is continuous. For $x\in \overline{\bb}$ we get
	\begin{align}
		\lim_{\substack{\tau \to 0 \\ \tau >0}} \int_{\mathcal{B}} \left( p^{\St}_i(x-y)-p^{\St}_{i,\tau}(x-y)\right)f_i(y)  \ \dy  =0,\quad i=1,2.
	\end{align}

	\begin{proof}
		Due to the construction, the difference of functions have the support $\mathbb{B}_\tau (x)$ for sufficiently small $\tau$.
		With the triangle inequality and dragging the $f$-component outside of the integral with its maximum, we obtain
		\begin{align}
			& \left| \int_{\mathbb{B}_\tau (x) \cap \overline{\mathcal{B}}} \left( p_i^{\St}(x-y)-p^{\St}_{i,\tau}(x-y)\right)f_i(y)  \ \dy \right| \notag\\
			& \leq \max_{y \in \mathbb{B}_\tau (x)\cap \overline{\mathcal{B}}} \left| f_i(y) \right| \int_{\mathbb{B}_\tau (x)\cap \overline{\mathcal{B}}} \left| p_i^{\St}(x-y)-p^{\St}_{i,\tau}(x-y)  \right| \dy \notag \\
			& \leq \max_{y \in \mathbb{B}_\tau (x)\cap \overline{\mathcal{B}}} \left| f_i(y) \right| \int_{\mathbb{B}_\tau (0)} \left| p_i^{\St}(y)-p^{\St}_{i,\tau} (y)  \right| \dy, \quad i=1,2.
		\end{align}
	 	Due the symmetry of the functions, it is sufficient first to consider the first quadrant of the $y$-domain and second to show the proof only for the first component $p^\St_{1}-p^\St_{1,\tau}$ by using polar coordinates. With the help of \cref{theorem:L-L_tau}, we can drop out the absolute value.
		\begin{align}
			\lim_{\substack{\tau \to 0 \\ \tau >0}}  \int_{\substack{\mathbb{B}_\tau(0) \notag\\ y_1,y_2 >0}} & \left| p_1^{\St}(y)-p^{\St}_{1,\tau}(y)\right| \ \dy \\
			&= \frac{C_1}{2\pi}\lim_{\substack{\tau \to 0 \\ \tau >0}}  \int_{\substack{\mathbb{B}_\tau(0) \notag\\ y_1,y_2 >0}} \frac{y_1}{\y^2}-\frac{y_1}{\tau^2}\left( 2-\frac{\y^2}{\tau^2} \right) \ \dy  \notag\\
			&=\frac{C_1}{2\pi} \lim_{\substack{\tau \to 0 \\ \tau >0}} \int_0^\tau \int_0^{\frac{\pi}{2}} \left( \frac{r \cos \ph}{r^2}-\frac{r \cos \ph}{\tau^2} \left( 2-\frac{r^2}{\tau^2} \right)\right) r \ \mathrm{d} \ph \ \mathrm{d} r \notag\\
			&= \frac{C_1}{2\pi}\lim_{\substack{\tau \to 0 \\ \tau >0}} \int_0^\tau \int_0^{\frac{\pi}{2}} \left( \cos \ph -2\frac{r^2 \cos \ph}{\tau^2}+\frac{r^4 \cos \ph}{\tau^4} \right) \ \mathrm{d} \ph \ \mathrm{d} r\notag\\
			&= \frac{C_1}{2\pi}\lim_{\substack{\tau \to 0 \\ \tau >0}} \int_0^\tau \left( 1-2\frac{r^2}{\tau^2}+\frac{r^4}{\tau^4} \right) \ \mathrm{d} r\notag\\
			&= \frac{C_1}{2\pi}\lim_{\substack{\tau \to 0 \\ \tau >0}} \left( \tau-\frac{2}{3} \frac{\tau^3}{\tau^2}+\frac{1}{5} \frac{\tau^5}{\tau^4} \right)\notag\\
			&= 0,
		\end{align}
		which yields our desired result.
	\end{proof}
\end{theorem}
We continue with the fundamental solution $u^\CN$ and obtain with the mollifications of the functions $L$ and $N$ from above  
\begin{align}
	u_{12,\tau}^{\CN}(x)& \coloneqq \frac{C_3 C_4}{2\pi} x_1 x_2 \begin{cases}
		\frac{1}{\|x\|^2},\ \|x\|\geq \tau,\\
		\frac{2}{\tau^2}-\frac{\x^2}{\tau^4} ,\ \|x\|<\tau.\end{cases}\\
	u_{kk,\tau}^{\CN}(x)& \coloneqq  \begin{cases}
		\frac{C_3}{2 \pi} \left(- \ln(\|x\|)+C_4\frac{x_k^2}{\|x\|^2} \right),\ \|x\|\geq \tau,\\
		\frac{C_3}{2 \pi} \left( - \ln(\tau)-\frac{\x^2}{2\tau^2}+\frac{1}{2}+C_4 x_k^2 \left( \frac{2}{\tau^2}-\frac{\|x\|^2}{ \tau^4}\right) \right),\ \|x\|<\tau.
	\end{cases}
\end{align}
Furthermore, we can prove the weak convergence.
\begin{theorem}
	We have the same assumptions as above, that means $\mathcal{B}$ is a regular region in $\R^2$ and $f:\overline{\bb} \to \R^2$ is continuous. Let $x\in \overline{\bb}$, then we get
	\begin{align}
		\lim_{\substack{\tau \to 0 \\ \tau >0}}  \left| \int_{\bb} \left( u_{ik}^{\CN}(x-y)-u^{\CN}_{ik,\tau}(x-y)\right)f_i(y) \ \dy \right| =0
	\end{align}
	for all $i,k=1,2$.
	\begin{proof}
		For the proof of this theorem, we have a look at two of the four components of $\boldsymbol{u}^{\CN}$ and $\boldsymbol{u}^{\CN}_\tau$ because we can use again the symmetry property.
		Again we have the compact support $\B_{\tau}(x)$ for the difference of the functions for sufficiently small $\tau$. With the triangle inequality and the maximum of the data $f$ like above, we obtain
		\begin{align}
			& \left| \int_{\mathbb{B}_\tau (x) \cap \overline{\mathcal{B}}} \left( u^\CN_{ik}(x-y)-u^\CN_{ik,\tau}(x-y) \right) f_i(y)\ \dy \right|\notag\\
			& \leq \max_{y \in \mathbb{B}_\tau (x) \cap \overline{\mathcal{B}}} \left| f_i(y) \right| \int_{\mathbb{B}_\tau (0)} \left| u^\CN_{ik}(y)-u^\CN_{ik,\tau}(y) \right|\ \dy, \quad i,k=1,2.
		\end{align}
		With the help of polar coordinates, we have a look at the several components separately. Again the absolute value can be dropped out due to \cref{theorem:L-L_tau} and the fact that $C_3,C_4$ are positive as a combination of positive material constants (see \eqref{eq:C3C4}). We get for the difference
		\begin{align}
			&\lim_{\substack{\tau \to 0 \\ \tau >0}} \int_{\mathbb{B}_\tau(0)} \left( u_{11}^{\CN}(y) -u_{11,\tau}^{\CN}(y) \right) \ \dy \notag\\
			&=\lim_{\substack{\tau \to 0 \\ \tau >0}} \int_{\mathbb{B}_\tau(0)}\frac{C_3}{2\pi} \left( -\ln \|y\|+C_4 \frac{y_1^2}{\|y\|^2}+\ln \tau +\frac{\|y\|^2}{2\tau^2} -\frac{1}{2}-C_4 y_1^2 \left( \frac{2}{\tau^2}-\frac{\|y\|^2}{\tau^4} \right) \right) \ \dy.
		\end{align}
		We split the integral and have a look at two separate terms
		\begin{align}
			\lim_{\substack{\tau \to 0 \\ \tau >0}}C_4 \int_{\mathbb{B}_\tau(0)}\left( \frac{y_1^2}{\|y\|^2}-y_1^2  \left( \frac{2}{\tau^2}-\frac{\|y\|^2}{\tau^4} \right) \right) \ \dy
			&=C_4 \lim_{\substack{\tau \to 0 \\ \tau >0}} \int_0^\tau \int_0^{2\pi} \left( \cos^2 \ph-r^2\cos^2 \ph \left( \frac{2}{\tau^2}-\frac{r^2}{\tau^4} \right) \right) r \ \mathrm{d}r \ \mathrm{d}\ph \notag \\
			&=C_4 \lim_{\substack{\tau \to 0 \\ \tau >0}} \left(\frac{1}{2}\tau^2 \cdot \frac{1}{2} \cdot 2\pi - \frac{1}{2} \cdot 2\pi \cdot \int_0^\tau \frac{2r^3}{\tau^2}-\frac{r^5}{\tau^4}\ \mathrm{d} r \right)\notag\\
			&=C_4\lim_{\substack{\tau \to 0 \\ \tau >0}} \left( \frac{1}{2} \tau^2 \pi- \pi \cdot \left( \frac{1}{2}\frac{\tau^4}{\tau^2}-\frac{1}{6} \frac{\tau^6}{\tau^4} \right) \right) \notag\\
			&=0.
		\end{align}
		The second part with the $\ln$-function is left:
				\begin{align}
			\lim_{\substack{\tau \to 0 \\ \tau >0}} \int_{\mathbb{B}_\tau(0)} -\ln \|y\|+\ln \tau+\frac{\|y\|^2}{2\tau^2}-\frac{1}{2} \ \dy
			\notag
			&= \lim_{\substack{\tau \to 0 \\ \tau >0}} \int_0^\tau \int_0^{2\pi} \left(-\ln r+\ln \tau+\frac{r^2}{2\tau^2}-\frac{1}{2}\right) r \ \mathrm{d}\ph\ \mathrm{d}r \notag\\
			&= \lim_{\substack{\tau \to 0 \\ \tau >0}} \left( 2\pi \int_0^\tau  -\ln r \cdot r+\ln\tau \cdot r+\frac{r^3}{2\tau^2}-\frac{1}{2}r \ \mathrm{d}r \right)\notag\\
			&= \lim_{\substack{\tau \to 0 \\ \tau >0}} 2\pi \left( -\frac{1}{4}r^2(2 \ln  r-1)+\frac{1}{2} r^2 \ln \tau+\frac{1}{4} \frac{r^4}{2\tau^2}-\frac{1}{4}r^2 \right)\Bigg|_0^\tau\notag\\
			&= \lim_{\substack{\tau \to 0 \\ \tau >0}} 2\pi \left( -\frac{1}{4}\tau^2 (2 \ln \tau-1)+\frac{1}{2} \tau^2 \ln \tau+\frac{1}{4} \frac{\tau^4}{2\tau^2}-\frac{1}{4}\tau^2 \right) \notag\\
			&= 0,
		\end{align}
		where $\lim_{b \to 0, b>0} b^2\ln b=0$ due to l'Hospital's rule.
		The function $u_{12,\tau}^\CN$ is left. Again due to the symmetry, we only consider the function for $y_1,y_2 >0$ and obtain subsequently with polar coordinates
				\begin{align}
			\lim_{\substack{\tau \to 0 \\ \tau >0}} \int_{\substack{\mathbb{B}_\tau(0) \\ y_1,y_2>0}} \left( u_{12}^{\CN}(y) -u_{12,\tau}^{\CN}(y) \right) \ \dy \notag
			&=\lim_{\substack{\tau \to 0 \\ \tau >0}} \frac{C_3 C_4}{2\pi} \int_{\substack{\mathbb{B}_\tau(0) \\ y_1,y_2>0}} \frac{y_1 y_2}{\|y\|^2}-y_1y_2 \left( \frac{2}{\tau^2}-\frac{\|y\|^2}{\tau^4} \right) \ \dy\notag \\
			&=\lim_{\substack{\tau \to 0 \\ \tau >0}} \frac{C_3 C_4}{2\pi} \int_0^\tau  \int_0^{\pi/2} \frac{r^2 \cos \ph \sin \ph}{r^2}\cdot r-r^2 \cos \ph \sin \ph \left(\frac{2}{\tau^2}-\frac{r^2}{\tau^4}\right)  \cdot r \ \mathrm{d} \ph \ \mathrm{d}r \notag\\
			&=\lim_{\substack{\tau \to 0 \\ \tau >0}}\frac{C_3 C_4}{2\pi} \left( \int_0^{\pi/2} \sin \ph \cos \ph \ \mathrm{d}\ph \cdot \int_0^\tau r-\frac{2r^3}{\tau^2}+\frac{r^5}{\tau^4} \ \mathrm{d}r\right) \notag\\
			&=\lim_{\substack{\tau \to 0 \\ \tau >0}} \frac{C_3 C_4}{2\pi}\left( \frac{1}{2} \cdot \left( \frac{1}{2}\tau^2-\frac{1}{2} \tau^2+\frac{1}{6} \tau^2 \right)   \right)\notag\\
			&=0.\tag*{\qedhere}
		\end{align}
	\end{proof}
\end{theorem}
We go over to the functions containing a space and time dependency. The approach is similar to the one above. We only regularize the spatial part of the functions and leave the time dependency as it is.
We can show with the concrete sequences $x^k=(1/k,1/k)$ and $t^k=1/k$ that the limit does not exist for $x=(0,0)$ and $t=0$ by inserting them in $p^\Si$
\begin{align}
	\lim_{k\to \infty} \frac{1}{4 \pi \frac{1}{k}} \exp\left( -\frac{\frac{2}{k^2}}{4C_2 \frac{1}{k}} \right)=\lim_{k\to \infty} \frac{k}{4\pi} \exp \left( -\frac{1}{2C_2k} \right)=\infty.
\end{align}
That means we get the mollification for $p^\Si$ by following the same steps as above
\begin{align}
	p^{\Si}_{\tau}(x,t)\coloneqq\begin{cases}
		\frac{1}{4\pi t} \exp \left( -\frac{\|x\|^2}{4 C_2 t} \right), \ \|x\|\geq \tau\\
		\frac{1}{4\pi t} \exp \left(-\frac{\tau^2}{4 C_2 t}  \right) \left[1-\frac{1}{4 C_2 t} \left( \|x\|^2-\tau^2 \right)  \right], \ \|x\| < \tau.
	\end{cases}
\end{align}

It follows that the limit from above now exists for the mollified fundamental solution by observing
\begin{align}
	\lim_{t\to 0+} \lim_{\|x\| \to 0} p^\Si_\tau(x,t) &= \lim_{t\to 0+} \lim_{\|x\| \to 0} \left( \frac{1}{4\pi t}\exp\left(-\frac{\tau^2}{4C_2 t} \right) \left[ 1-\frac{1}{4C_2t} (\|x\|^2-\tau^2)  \right] \right)\notag \\
	&=\lim_{t\to 0+} \lim_{\|x\| \to 0} \left[ \frac{1}{4\pi t}\exp\left(-\frac{\tau^2}{4C_2 t}\right) \left( 1+\frac{\tau^2}{4C_2t}  \right) \right] \notag \\
	&\phantom{=}-\lim_{t\to 0+} \lim_{\|x\| \to 0} \frac{1}{4\pi t}\exp\left(-\frac{\tau^2}{4C_2 t}\right) \frac{\|x\|^2}{4C_2t} \notag \\
	&=\lim_{t\to 0+}  \left[ \frac{1}{4\pi t}\exp\left(-\frac{\tau^2}{4C_2 t}\right) \left( 1+\frac{\tau^2}{4C_2t}  \right) \right] \notag \\
	&\phantom{=}-\lim_{t\to 0+} \frac{1}{16C_2 \pi t^2}\exp\left(-\frac{\tau^2}{4C_2 t}\right)  \cdot \lim_{\|x\| \to 0} \|x\|^2 \notag \\
	&=0.
\end{align}
In analogy to above, we can show the following theorem
\begin{theorem}
	Let $\mathcal{B}$ a regular region in $\R^2$ and $f:\overline{\bb} \times \R \to \R$ be continuous and bounded. With $x\in \overline{\bb}$ and $t\in \R^+$, we get
	\begin{align}
		\lim_{\substack{\tau \to 0 \\ \tau >0}} \left| \int_{t-T}^{t} \int_{\mathcal{B}} \left( p^\Si(x-y,t-\theta)-p^{\Si}_\tau(x-y,t-\theta)\right)f(y,\theta)  \ \dy \ \mathrm{d} \theta \right| =0,
	\end{align}
	where $T>0$ is the length of our considered time interval.
	\begin{proof}
		We follow similar steps as in the proofs above. That means first using the triangle inequality and dragging the $f$-component outside of the integral with its supremum.
		Here the spatial support for a fixed $t$ of the difference of functions is $\mathbb{B}_\tau (x)$ for sufficiently small $\tau$. We obtain
		\begin{align}
			& \left| \int_{t-T}^{t} \int_{\mathbb{B}_\tau (x) \cap \overline{\mathcal{B}}} \left( p^\Si(x-y,t-\theta)-p^{\Si}_\tau(x-y,t-\theta)\right)f(y,\theta)  \ \dy \ \mathrm{d} \theta \right| \notag\\
			& \leq \sup_{y \in \mathbb{B}_\tau (x) \cap \overline{\mathcal{B}}, \theta \in \R} \left| f(y,\theta) \right| \int_{0}^{T} \int_{\mathbb{B}_\tau (0)} \left| p^{\Si}(y,\theta)-p^{\Si}_\tau(y,\theta)  \right| \dy \ \mathrm{d}\theta.
		\end{align}
		Note that we assumed that $f$ is bounded.
		Now we get that $p^\Si-p^\Si_\tau \geq 0$ holds true by proving that this difference is monotonically decreasing with respect to $\|x\|=r$, where we assume a positive time $t>0$:
		\begin{align}
			\frac{\partial}{\partial r} \left( p^\Si-p_\tau^\Si \right)(x,t)&= \frac{\partial}{\partial r} \frac{1}{4\pi t} \left[\exp \left( \frac{-r^2}{4 C_2 t} \right)-\exp \left( \frac{-\tau^2}{4 C_2 t} \right)\left( 1-\frac{1}{4C_2t}(r^2-\tau^2) \right)  \right] \notag\\
			&= \frac{1}{4\pi t} \left[\exp \left( \frac{-r^2}{4 C_2 t} \right) \cdot \left( \frac{-2r}{4C_2 t} \right)-\exp \left( \frac{-\tau^2}{4 C_2 t} \right) \cdot \left( \frac{-1}{4C_2 t}\right) \cdot 2r  \right]\notag\\
			&= \underbrace{\frac{1}{4\pi t}}_{>0} \cdot \underbrace{\left( \frac{-2r}{4C_2 t} \right)}_{<0} \underbrace{\left[ \exp \left( \frac{-r^2}{4 C_2 t} \right)-\exp \left( \frac{-\tau^2}{4 C_2 t}\right)  \right]}_{>0\ ( \text{for } r<\tau)} <0.
		\end{align}
		Now we have that the difference is monotonically decreasing for $\|x\|=r$ and together with the zero value at $\|x\|=\tau$, we obtain that the difference is non-negative for all $t>0$ because our consideration and estimation was independent of $t$. Due to the construction with the Taylor expansion, we have that $p^\Si_\tau \in \mathrm{C}^{(1)}(\R^2 \times \R)$. 
		That means we can drop out the absolute value of the integral and get with polar coordinates (see also
		\cref{remark:exp_integral} for the time integrals)
		\begin{align}
			&\int_0^T \int_{\B_\tau (0)} \frac{1}{4\pi t} \exp \left( \frac{-\|x\|^2}{4 C_2 t} \right)-\frac{1}{4\pi t} \exp \left( \frac{-\tau^2}{4 C_2 t} \right) \left[ 1- \frac{1}{4C_2 t} (\|x\|^2-\tau^2) \right] \ \dx \ \mathrm{d}t\notag\\
			&= \int_0^T \int_0^{2\pi} \int_0^\tau \frac{r}{4\pi t} \exp \left( \frac{-r^2}{4 C_2 t} \right) -\frac{r}{4\pi t} \exp \left( \frac{-\tau^2}{4 C_2 t} \right) \left[ 1-\frac{1}{4C_2t} (r^2-\tau^2) \right] \ \mathrm{d}r \ \mathrm{d} \ph \ \mathrm{d} t \notag\\
			&= 2\pi \int_0^T \int_0^\tau \frac{r}{4\pi t} \exp \left( \frac{-r^2}{4 C_2 t} \right) -\frac{1}{4\pi t} \exp \left( \frac{-\tau^2}{4 C_2 t} \right) \left[r-\frac{r^3}{4 C_2 t}+\frac{r\tau^2}{4 C_2 t} \right] \ \mathrm{d} r \ \mathrm{d} t\notag\\
			&= 2\pi \cdot \frac{1}{2\pi} \int_0^T -C_2  \exp \left( \frac{-\tau^2}{4 C_2 t} \right)+C_2-\frac{1}{2 t} \exp \left( \frac{-\tau^2}{4 C_2 t} \right) \left[ \frac{1}{2}\tau^2-\frac{\tau^4}{16 C_2 t}+\frac{\tau^4}{8C_2 t} \right] \ \mathrm{d}t\notag\\
			&=  C_2 \left(T-\frac{\tau^2 \Ei \left(-\frac{\tau^2}{4C_2 T}\right) }{4C_2}-T \exp \left( \frac{-\tau^2}{4 C_2 T} \right) \right)\notag\\
			&\phantom{=} +\Ei \left(-\frac{\tau^2}{4C_2 T}\right)\cdot \frac{\tau^2}{4}-\frac{\tau^4}{32 C_2}\cdot \frac{4C_2 \exp \left(-\frac{\tau^2}{4C_2 T}\right)}{\tau^2}  \notag\\
			&\to \ 0\quad (\text{as }\tau \to 0+),
		\end{align}
		where we considered \eqref{eq:Eilimes}. This completes our proof.
	\end{proof}
\end{theorem}

The last function to consider is $u^\Si$. Here again the limit for $t=0$ and $x=(0,0)$ does not exist, which
we can show with the help of the sequences $x^k=(1/k,1/k)$ and $t^k=1/k^2$. This gives us
\begin{align}
	\lim_{k\to \infty} \frac{C_1}{2\pi} \frac{k}{2} \left( 1-\underbrace{\exp \left( -\frac{2/k^2}{4C_2 \cdot  1/k^2} \right)}_{= \mathrm{const.} < 1} \right)=\infty.
\end{align}

Following the same principle as in the case of $p^\Si$, that means only applying the Taylor expansion to the spatial part, and ignoring the $x$-term like in the case of $u^\CN$, we get
\begin{align}
	u^{\Si}_{\tau}(x,t)=
	\begin{cases}
		&\frac{C_1}{2\pi} \frac{x}{ \|x\|^2} \left( 1-\exp \left( - \frac{\|x\|^2}{4 C_2 t} \right) \right),\ \|x\| \geq \tau \\
		& \frac{C_1 }{2 \pi} x \left[ \frac{1}{\tau^2}-\frac{1}{\tau^2} \exp \left( -\frac{\tau^2}{4 C_2 t} \right) +\left( \frac{1}{\tau^2}\frac{1}{4 C_2 t} \exp \left( -\frac{\tau^2}{4 C_2 t} \right)  \right. \right. \\
		&\left. \left. -  \frac{1}{ \tau^4}+\frac{1}{ \tau^4} \exp \left(-\frac{\tau^2}{4 C_2 t}  \right) \right)(\|x\|^2-\tau^2) \right],\ \|x\| <\tau.
	\end{cases}
\end{align}

Again, we can show that the limit exists in the case of the mollified fundamental solution \begin{align}
	&\lim_{t\to 0+} \lim_{\|x\| \to 0} u^\Si_{1,\tau} (x,t)\notag \\
	&= \lim_{t\to 0+} \lim_{\|x\| \to 0}\frac{C_1 }{2 \pi} x_1 \left[ \frac{1}{\tau^2}\left( 1- \exp \left( -\frac{\tau^2}{4 C_2 t} \right)\right) \right.\notag \\
	&\phantom{=}\left.+ \left( \frac{1}{\tau^2}\frac{1}{4 C_2 t} \exp \left( -\frac{\tau^2}{4 C_2 t} \right)+\frac{1}{ \tau^4}\left( \exp \left(-\frac{\tau^2}{4 C_2 t}  \right)-1\right) \right)(\|x\|^2-\tau^2) \right]\notag \\
	&=  \lim_{t\to 0+} \lim_{\|x\| \to 0}\frac{C_1 }{2 \pi}x_1 \left[ \frac{1}{\tau^2}\left( 1- \exp \left( -\frac{\tau^2}{4 C_2 t} \right)\right) \right.\notag \\
	&\phantom{=}\left.+ \left(-\frac{1}{4 C_2 t} \exp \left( -\frac{\tau^2}{4 C_2 t} \right)-\frac{1}{ \tau^2}\left( \exp \left(-\frac{\tau^2}{4 C_2 t}  \right)-1\right) \right) \right]\notag \\
	&\phantom{=}+ \lim_{t\to 0+} \lim_{\|x\| \to 0} \frac{C_1}{2\pi} x_1 \|x\|^2 \left( \frac{1}{\tau^2}\frac{1}{4 C_2 t} \exp \left( -\frac{\tau^2}{4 C_2 t} \right)+\frac{1}{ \tau^4}\left( \exp \left(-\frac{\tau^2}{4 C_2 t}  \right)-1\right) \right) \notag \\
	&= \frac{C_1}{2\pi} \cdot 0 \cdot \frac{2}{\tau^2}+\frac{C_1}{2\pi} \cdot 0 \cdot \left(-\frac{1}{\tau^4}  \right) \notag \\
	&=0
\end{align}
and we can also prove the weak convergence.
\begin{theorem}
	Let $\mathcal{B}$ be a regular region in $\R^2$ and $f:\overline{\bb} \times \R \to \R^2$ be continuous and bounded. With $x\in \overline{\bb}$ and $t\in \R^+$, we get
	\begin{align}
		\lim_{\substack{\tau \to 0 \\ \tau >0}} \left| \int_{t-T}^{t} \int_{\mathcal{B}} \left( u_i^\Si(x-y,t-\theta)-u^{\Si}_{i,\tau}(x-y,t-\theta)\right)f_i(y,\theta)  \ \dy \ \mathrm{d} \theta \right| =0,\quad i=1,2,
	\end{align}
	where $T>0$ is the length of our considered time interval.
	\begin{proof}	
		The steps are the same as for the $p^\Si$-part, i.e. using the triangle inequality and dragging the $f$-component outside of the integral with its supremum.
		The support of the spatial part is again $\mathbb{B}_\tau (x)$ with the same arguments as above. We get
		\begin{align}
			& \left| \int_{t-T}^{t} \int_{\mathbb{B}_\tau (x) \cap \overline{\mathcal{B}}} \left( u^\Si_i(x-y,t-\theta)-u^{\Si}_{i,\tau}(x-y,t-\theta)\right)f_i(y,\theta)  \ \dy \ \mathrm{d} \theta \right|\notag \\
			& \leq \sup_{y \in \mathbb{B}_\tau (x) \cap \overline{\mathcal{B}}, \theta \in \R} \left| f_i(y,\theta) \right| \int_0^T \int_{\mathbb{B}_\tau (0)} \left| u^{\Si}_i(y,\theta)-u^{\Si}_{i,\tau}(y,\theta)  \right| \dy \ \mathrm{d}\theta.
		\end{align}
		We show the next steps of this theorem without loss of generality only for the first component, the second one is obtained due to the symmetry.
		First, we neglect the factor $\frac{C_1}{2\pi} x_1$, consider the difference of $u^\Si_1-u^\Si_{1,\tau}$ for $\|x\|<\tau$ and get
		\begin{align}
			&\frac{1}{\|x\|^2} \left( 1-\exp\left( -\frac{\|x\|^2}{4C_2t} \right) \right)-\frac{1}{\tau^2} \left( 1-\exp\left( -\frac{\tau^2}{4C_2t} \right) \right)\notag \\
			&+\left( \frac{1}{\tau^2}\frac{1}{4C_2t} \exp\left( -\frac{\tau^2}{4C_2t} \right)+\frac{1}{\tau^4} \left( \exp\left( -\frac{\tau^2}{4C_2t} \right)-1 \right) \right) (\tau^2-\|x\|^2). \label{eq:uSiproofdiff}
		\end{align}
		Let us have a look at the two lines separately with the aim to show that the first one is non-negative and the second one is non-positive. We start our considerations with the second line. The term $(\tau^2-\|x\|^2)$ is positive since we are in the case $\|x\|<\tau$. The following conversions are helpful for us for the remaining first term
		\begin{align}
			&\frac{1}{\tau^2}\frac{1}{4C_2t} \exp\left( -\frac{\tau^2}{4C_2t} \right)+\frac{1}{\tau^4} \left( \exp\left( -\frac{\tau^2}{4C_2t} \right)-1 \right) \notag \\
			&= \frac{\tau^2 \exp\left( -\frac{\tau^2}{4C_2t} \right)+4C_2t \exp\left( -\frac{\tau^2}{4C_2t} \right)-4C_2t}{4C_2 t \tau^4}. \label{eq:uSiproof_1}
		\end{align}
		Due to the positivity of $\tau^4,t,C_2$ , the denominator is non-negative and we want to take a closer look at the numerator. It is differentiated with respect to $\tau$
		\begin{align}
			& 2 \tau \exp\left( -\frac{\tau^2}{4C_2t} \right)+\tau^2 \exp\left( -\frac{\tau^2}{4C_2t} \right) \cdot \left( -\frac{2\tau}{4C_2t} \right)+4C_2t \exp\left( -\frac{\tau^2}{4C_2t} \right) \cdot \left(-\frac{2\tau}{4C_2t}  \right) \notag\\
			&= \exp\left( -\frac{\tau^2}{4C_2t} \right) \cdot \left(  -\frac{2\tau^3}{4C_2t}\right) <0,
		\end{align}
		that means the numerator of \eqref{eq:uSiproof_1} is monotonically decreasing with respect to $\tau$. With this we can estimate it with its maximum for $\tau=0$.
		This insertion leads us to
		\begin{align}
			\tau^2 \exp\left( -\frac{\tau^2}{4C_2t} \right)+4C_2t \exp\left( -\frac{\tau^2}{4C_2t} \right)-4C_2t \leq 0+4C_2t-4C_2t=0.
		\end{align}
		Thereby we know that \eqref{eq:uSiproof_1} is non-positive and therefore the second line of \eqref{eq:uSiproofdiff}, too. Let us have a look at the first line of \eqref{eq:uSiproofdiff} and set $c\coloneqq 4C_2t$ for the sake of readability. We can show that the function
		\begin{align}
			\frac{1}{r^2} \left(1-\exp\left( -\frac{r^2}{4C_2t} \right)  \right) \label{eq:uSiproof4}
		\end{align}
		is monotonically decreasing in $r$. Considering the derivative with respect to $r$ leads us to
		\begin{align}
			&-\frac{2}{r^3} \left( 1- \exp\left(-\frac{r^2}{c} \right) \right)+\frac{1}{r^2} \exp \left( -\frac{r^2}{c} \right) \frac{2r}{c}\notag  \\
			&= \frac{2}{cr^3}\left[ -c+c  \exp\left(-\frac{r^2}{c} \right) +r^2 \exp\left(-\frac{r^2}{c} \right)\right], \label{eq:uSiproof3}
		\end{align}
		for which we intend to show the non-positivity. Since the first factor is positive, we want to have a closer look at the term in the squared brackets. Our aim is to get the maximum of it, which we get by deriving it again with respect to $r$
		\begin{align}
			&c \exp\left(-\frac{r^2}{c} \right) \cdot \left(-\frac{2r}{c} \right)+2r \exp\left(-\frac{r^2}{c} \right)+r^2 \exp\left(-\frac{r^2}{c} \right) \cdot \left( -\frac{2r}{c} \right)\notag \\
			&= \exp\left(-\frac{r^2}{c} \right) \cdot \left( -\frac{2r^3}{c}\right). \label{eq:uSiproof2}
		\end{align}
		Possible roots are given by $r=0$ or $c=0$ (i.e. $c \to 0+$), whereas we want to have a look at the case $c=0$ later separately.
		In the case $r=0$, we can see directly from \eqref{eq:uSiproof2} that we have a conversion from positive to negative sign around $r=0$ which means that we have a maximum.
		With this, we can estimate the squared brackets in \eqref{eq:uSiproof3} by
		\begin{align}
			-c+c  \exp\left(-\frac{r^2}{c} \right) +r^2 \exp\left(-\frac{r^2}{c} \right) \leq -c+c+0=0
		\end{align}
		and obtain that \eqref{eq:uSiproof3} is non-positive and therefore the function in \eqref{eq:uSiproof4} is monotonically decreasing. Furthermore, the difference in the first line of \eqref{eq:uSiproofdiff} is non-negative. The case $c\to 0+$ reflects the case $t\to 0+$ and results in the function $1/r^2$ instead of \eqref{eq:uSiproof4}, which is obviously monotonically decreasing.
		Now we come back to the original integral above and estimate the absolute value of the difference with the triangle inequality. For the first term,  we can drop out the absolute value (except for the term $x_1$) since we showed that the first term is positive. We can omit the absolute value in the second by changing the sign in front of the term since the term in the absolute value is negative
		\begin{align}
			\left| u_1^\Si (x,t) -u^\Si_{1,\tau} (x,t) \right| \notag
			& \leq \frac{C_1}{2\pi} | x_1| \left| \frac{1}{\|x\|^2} \left( 1-\exp\left( -\frac{\|x\|^2}{4C_2t} \right) \right)-\frac{1}{\tau^2} \left( 1-\exp\left( -\frac{\tau^2}{4C_2t} \right) \right)  \right| \notag \\
			& \phantom{\leq} +  \frac{C_1}{2\pi} | x_1| \left| \left( \frac{1}{\tau^2}\frac{1}{4C_2t} \exp\left( -\frac{\tau^2}{4C_2t} \right)+\frac{1}{\tau^4} \left( \exp\left( -\frac{\tau^2}{4C_2t} \right)-1 \right) \right) \right| \left| \tau^2-\|x\|^2  \right| \notag \\
			& \leq \frac{C_1}{2\pi} | x_1| \left( \frac{1}{\|x\|^2} \left( 1-\exp\left( -\frac{\|x\|^2}{4C_2t} \right) \right)-\frac{1}{\tau^2} \left( 1-\exp\left( -\frac{\tau^2}{4C_2t} \right) \right)  \right) \notag \\
			& \phantom{\leq} -  \frac{C_1}{2\pi} | x_1| \left(  \frac{1}{\tau^2}\frac{1}{4C_2t} \exp\left( -\frac{\tau^2}{4C_2t} \right)+\frac{1}{\tau^4} \left( \exp\left( -\frac{\tau^2}{4C_2t} \right)-1 \right)  \right) \left( \tau^2-\|x\|^2  \right)
		\end{align}
		Now we consider the integral above without the absolute value and get with polar coordinates for the case $x_1>0$:
		\begin{align}
			&\frac{C_1}{2\pi} \int_0^T \int_{\substack{\B_\tau (0), \\ x_1>0}} \frac{x_1}{\|x\|^2} \left( 1-\exp \left(-\frac{\|x\|^2}{4C_2 t}\right)\right)\notag\\
			&\phantom{=}-x_1 \left[ \frac{1}{\tau^2} \left( 1-\exp \left(-\frac{\tau^2}{4C_2 t}\right)\right) \right.\notag\\
			&\phantom{=}\hspace*{1cm}\left.  - \left( -\frac{1}{\tau^4} \left( 1-\exp \left(-\frac{\tau^2}{4C_2 t}\right)\right) + \frac{1}{\tau^2}\frac{1}{4C_2 t}\exp \left(-\frac{\tau^2}{4C_2 t}\right) \right)\cdot (\|x\|^2-\tau^2) \right] \ \dx \ \mathrm{d}t\notag\\
			&= \frac{C_1}{2\pi} \int_{-\pi/2}^{\pi/2} \cos \ph \ \mathrm{d}\ph \notag\\
			&\phantom{=}\times \int_0^T \int_0^\tau \frac{r^2}{r^2}\left( 1-\exp \left(-\frac{r^2}{4C_2 t}\right)\right)\notag\\
			&\phantom{=}- \left[\frac{r^2}{\tau^2} \left( 1-\exp \left(-\frac{\tau^2}{4C_2 t}\right)\right)\right.\notag\\
			&\qquad \phantom{=}\left. -\left( -\frac{1}{\tau^4} \left( 1-\exp \left(-\frac{\tau^2}{4C_2 t}\right)\right) +\frac{1}{\tau^2}\frac{1}{4 C_2t} \exp \left(-\frac{\tau^2}{4C_2 t}\right)\right) (r^4-r^2 \tau^2) \right]\ \mathrm{d}r \ \mathrm{d}t. \label{eq:u-uSi}
		\end{align}
		The first integral gives us
		\begin{align}
			\int_{-\pi/2}^{\pi/2} \cos \ph \ \mathrm{d}\ph=2.
		\end{align}
		We have a look at the second line of the integral separately later. First we come to the third and fourth line of the integral, which is integrated with respect to $r$ and evaluated (see also \cref{remark:exp_integral}) afterwards. We get
		\begin{align}
			&\int_0^T -\frac{1}{3} \frac{\tau^3}{\tau^2}\left( 1-\exp \left(-\frac{\tau^2}{4C_2 t}\right)\right)\notag\\
			&\phantom{=}+ \left( \frac{1}{\tau^4}\left( 1-\exp \left(-\frac{\tau^2}{4C_2 t}\right)\right)-\frac{1}{\tau^2}\frac{1}{4C_2t}\exp \left(-\frac{\tau^2}{4C_2 t}\right) \right) \frac{2}{15} \tau^5 \ \mathrm{d}t\notag\\
			&= -\frac{1}{3} \tau \left( T-\frac{\tau^2 \Ei\left(-\frac{\tau^2}{4C_2 T}\right)}{4C_2}-T \exp \left(-\frac{\tau^2}{4C_2 T}\right) \right)\notag\\
			&\phantom{=}+\left( \frac{1}{\tau^4} \left( T-\frac{\tau^2 \Ei\left(-\frac{\tau^2}{4C_2 T}\right)}{4C_2}-T \exp \left(-\frac{\tau^2}{4C_2 T}\right) \right) \right. \notag\\
			&\quad \phantom{=-}\left. -\frac{1}{\tau^2} \frac{1}{4C_2} \cdot \left(- \Ei\left(-\frac{\tau^2}{4C_2 T}\right) \right) \vphantom{\frac{\tau^2 \Ei\left(-\frac{\tau^2}{4C_2 T}\right)}{4C_2}}  \right) \frac{2}{15} \tau^5\notag\\
			&\to 0\ (\tau \to 0+)
		\end{align}
		with the help of the considerations from \eqref{eq:Eilimes}.
		Now we have a look at the second line of the integral in \eqref{eq:u-uSi}, which is
		\begin{align}
			\int_0^T \int_0^\tau 1-\exp\left(-\frac{r^2}{4C_2 t}\right)\ \mathrm{d} r \ \mathrm{d} t \notag
			&	= \int_0^T \tau-\frac{1}{2}\sqrt{\pi} \sqrt{4C_2 t}\ \mathrm{erf} \left(\frac{\tau}{\sqrt{4C_2 t}}\right) \ \mathrm{d}t.
		\end{align}
		The first part with $\tau$ is convergent to $0$ for $\tau \to 0+$.
		For the second part we obtain with the help of \eqref{eq:Eilimes}
		\begin{align}
			\int_0^T \sqrt{4C_2 t} \ \mathrm{erf} \left(\frac{\tau}{\sqrt{4C_2 t}}\right) \ \mathrm{d}t \notag
			&= \frac{2}{3} \left( \frac{\tau \left( \frac{\tau^2 \Ei \left(-\frac{\tau^2}{4C_2 T}\right)}{4C_2}+T \exp\left(-\frac{\tau^2}{4C_2 T}\right) \right)}{\sqrt{\pi}}+T\cdot \sqrt{4C_2 T} \mathrm{erf} \left(\frac{\tau}{\sqrt{4C_2 T}}\right) \right)\notag\\
			& \to 0 \ (\tau \to 0+).
		\end{align}
		This primitive function can easily be verified by deriving (see also \cref{def:Ei} and the definition of the error function)
		\begin{align}
			&\frac{\partial }{\partial t} \left[ \frac{2}{3} \left( \frac{\tau \left( \frac{\tau^2 \Ei \left(-\frac{\tau^2}{4C_2 t}\right)}{4C_2}+t \exp\left(-\frac{\tau^2}{4C_2 t}\right) \right)}{\sqrt{\pi}}+t\cdot \sqrt{4C_2 t}\ \mathrm{erf} \left(\frac{\tau}{\sqrt{4C_2 t}}\right) \right)  \right] \notag \\
			&=\frac{2}{3} \left[ \frac{\tau}{\sqrt{\pi}} \left( \frac{\tau^2}{4C_2} \exp\left(-\frac{\tau^2}{4C_2 t}\right) \cdot \left( \frac{-4C_2 t}{\tau^2}  \right) \cdot \left( \frac{\tau^2}{4C_2 t^2} \right) +\exp \left(-\frac{\tau^2}{4C_2 t}\right)  \right.\right.\notag\\
			&\phantom{=}\left. \left.+t \exp\left(-\frac{\tau^2}{4C_2 t}\right) \cdot \frac{\tau^2}{4C_2t^2}\right) +\sqrt{4C_2t}\ \mathrm{erf} \left(\frac{\tau}{\sqrt{4C_2 t}}\right)+t \frac{4C_2}{2 \sqrt{4C_2t}}\mathrm{erf} \left(\frac{\tau}{\sqrt{4C_2 t}}\right)\right.\notag\\
			&\phantom{=} \left. +t\sqrt{4 C_2 t}\frac{2}{\sqrt{\pi}}  \exp\left(-\frac{\tau^2}{4C_2 t}\right) \cdot \left(-\frac{\tau \cdot 4 C_2}{2 (4 C_2 t)^{3/2}} \right)   \right] \notag\\
			&= \frac{2}{3} \left[ \frac{\tau}{\sqrt{\pi}} \left(  -\frac{\tau^2}{4C_2t}  \exp\left(-\frac{\tau^2}{4C_2 t}\right)  + \exp\left(-\frac{\tau^2}{4C_2 t}\right) +\frac{\tau^2}{4C_2t}\exp\left(-\frac{\tau^2}{4C_2 t}\right) \right) \right. \notag\\
			&\phantom{=}\left.+ \sqrt{4C_2t}\ \mathrm{erf} \left(\frac{\tau}{\sqrt{4C_2 t}}\right)+\sqrt{C_2t}\ \mathrm{erf} \left(\frac{\tau}{\sqrt{4C_2 t}}\right) -\frac{\tau}{\sqrt{\pi}}  \exp\left(-\frac{\tau^2}{4C_2 t}\right)\right] \notag \\
			&=\frac{2}{3} \mathrm{erf} \left(\frac{\tau}{\sqrt{4C_2 t}}\right) \left( \sqrt{4C_2t}+\frac{1}{2} \sqrt{4C_2t} \right) \notag\\
			&=\sqrt{4C_2t}\ \mathrm{erf} \left(\frac{\tau}{\sqrt{4C_2 t}}\right).
		\end{align}
		With this last proof, we finished the proofs for all weak convergence results for our mollified fundamental solutions.
	\end{proof}	
\end{theorem}

\section{Source scaling functions}
\subsection{Derivation}
For the construction of the source scaling functions, we need the differential operator $L^\pe$ given by
\begin{align}
	L^{\pe}(u,p)= \begin{pmatrix}
		-\frac{\lambda+\mu}{\mu} \nabla_x (\nabla_x \cdot u)-\nabla_x^2 u+\alpha \nabla_x p \\
		\partial_t (c_0 \mu p+\alpha( \nabla_x \cdot u))-\nabla_x^2 p \label{eq:differential_operator}
	\end{pmatrix},
\end{align}

and apply it now on $G_\tau$, which has the regularized fundamental solutions from above as entries
\begin{align}
	\boldsymbol{G}_{\tau}(x,t)=\begin{pmatrix}
		u_{11,\tau}^{\CN}(x)\delta_t & u_{12,\tau}^{\CN}(x)\delta_t & p_{1,\tau}^{\St}(x)\delta_t \\
		u_{21,\tau}^{\CN}(x)\delta_t & u_{22,\tau}^{\CN}(x)\delta_t &  p_{2,\tau}^{\St}(x) \delta_t \\
		u_{1,\tau}^{\Si}(x,t) &u_{2,\tau}^{\Si}(x,t) & p_\tau^{\Si}(x,t)
	\end{pmatrix}.
\end{align}
We do this by applying the differential operator on each row of $G_\tau$  (see \eqref{def:divergence}).
The entries of the resulting tensor are called the source scaling functions and are denoted by
\begin{align}
	\boldsymbol{\Phi}_{\tau}(x,t)=\begin{pmatrix}
		\Phi_{11,\tau}(x)\delta_t & \Phi_{12,\tau}(x)\delta_t &\Phi_{13,\tau}^{1}(x)\delta'_t+\Phi_{13,\tau}^{2}(x) \delta_t \\
		\Phi_{21,\tau}(x)\delta_t & \Phi_{22,\tau}(x)\delta_t & \Phi_{23,\tau}^{1}(x)\delta'_t+\Phi_{23,\tau}^{2}(x)\delta_t\\
		\Phi_{31,\tau}(x,t) & \Phi_{32,\tau}(x,t) & \Phi_{33,\tau}(x,t)
	\end{pmatrix} \label{eq:source_functions_tensor}
\end{align}
Now we can figure out the several components.

We are using the following notation: $\Phi_{ik,\tau}$ are the source scaling functions without the delta distribution (if they contain such one) and by $\left(\boldsymbol{\Phi}_{\tau}\right)_{ik}$ we denote the whole entry of the tensor, that means they can include a delta distribution (like in the case of the first two rows).

Now we have to remember how we constructed our regularized fundamental solutions: In the case $\|x\|\geq \tau$ we use the fundamental solutions themselves and for $\|x\|<\tau$ a Taylor mollification of the fundamental solutions. Applying the differential operator on $\boldsymbol{G}_\tau$ results in the following: the source scaling functions have a compact support for $\|x\| \leq \tau$ for the spatial part because outside it is zero. Thus the derivations for the source scaling functions are done for the case $\|x\| \leq \tau$.

The detailed calculations for all entries can be found in the appendix.

For reasons of clarity, we summarize the components of $\boldsymbol{\Phi}_\tau$ here at a glance. We can see the several symmetry relations of the source scaling functions, that means for example $\Phi_{11,\tau}$ and $\Phi_{22,\tau}$, $\Phi_{13,\tau}$ and $\Phi_{23,\tau}$ and at last $\Phi_{31,\tau}$ and $\Phi_{32,\tau}$. The source scaling functions are
\begin{align}
	\Phi_{11,\tau}(x) &= -\frac{\lambda+\mu}{\mu} \frac{C_3}{2\pi} \left( -\frac{1}{\tau^2}+C_4 \cdot \left( \frac{6}{\tau^2}-\frac{15x_1^2+5x_2^2}{\tau^4} \right) \right)\notag\\
	&\phantom{=}-\frac{C_3}{2\pi} \left( \frac{-2+4C_4}{\tau^2}+C_4 \cdot \left( -\frac{14x_1^2+2x_2^2}{\tau^4} \right)  \right)+\alpha \frac{C_1}{2\pi} \frac{1}{\tau^2} \left( 2-\frac{3x_1^2+x_2^2}{\tau^2} \right), \\
	\Phi_{12,\tau} (x)&=\frac{x_1x_2}{\tau^4 \pi} \left( \frac{\lambda+\mu}{\mu} \cdot 5 C_3 C_4  +6C_3 C_4-\alpha C_1 \right), \\
	\Phi_{13,\tau}^1(x) &= \left[ \frac{c_0 \mu C_1}{2\pi}\cdot \frac{x_1}{\tau^2}\left( 2-\frac{\|x\|^2}{\tau^2}\right)+\frac{\alpha C_3}{2\pi} \left( \frac{x_1 (6C_4-1)}{\tau^2}-5C_4 \cdot x_1\cdot \frac{\|x\|^2}{\tau^4} \right) \right],  \\
	\Phi_{13,\tau}^2(x) &=\frac{4 x_1 C_1}{\tau^3 \pi},\\
	\Phi_{21,\tau}(x) &=\frac{x_1x_2}{\tau^4 \pi} \left( \frac{\lambda+\mu}{\mu} \cdot 5 C_3 C_4  +6C_3 C_4-\alpha C_1 \right), \\
	\Phi_{22,\tau} (x)&= -\frac{\lambda+\mu}{\mu} \frac{C_3}{2\pi} \left( -\frac{1}{\tau^2}+C_4 \cdot \left( \frac{6}{\tau^2}-\frac{15x_2^2+5x_1^2}{\tau^4} \right) \right)\notag\\
	&\phantom{=}-\frac{C_3}{2\pi} \left( \frac{-2+4C_4}{\tau^2}+C_4 \cdot \left( -\frac{14x_2^2+2x_1^2}{\tau^4} \right)  \right)+\alpha \frac{C_1}{2\pi} \frac{1}{\tau^2} \left( 2-\frac{3x_2^2+x_1^2}{\tau^2} \right), \\
	\Phi_{23,\tau}^1(x) &= \left[ \frac{c_0 \mu C_1}{2\pi}\cdot \frac{x_2}{\tau^2}\left( 2-\frac{\|x\|^2}{\tau^2}\right)+\frac{\alpha C_3}{2\pi} \left( \frac{x_2 (6C_4-1)}{\tau^2}-5C_4 \cdot x_2\cdot \frac{\|x\|^2}{\tau^4} \right) \right],\\
	\Phi_{23,\tau}^2(x) &=\frac{4 x_2 C_1}{\tau^3 \pi},\\
	\Phi_{31,\tau} (x,t)&= -x_1 \exp \left( -\frac{\tau^2}{4 C_2 t} \right)  \left( \vphantom{\frac{\left(1-\exp \left( \frac{\tau^2}{4 C_2 t} \right)\right) }{8C_2 \mu t^2 \tau^4 \pi}} \frac{\alpha\mu \tau^4+8C_1(\lambda+2\mu)t\tau^2}{8C_2 \mu t^2 \tau^4 \pi} + \frac{32C_1C_2(\lambda+2\mu)t^2\left(1-\exp \left( \frac{\tau^2}{4 C_2 t} \right)  \right)}{8C_2 \mu t^2 \tau^4 \pi}   \right) , \\
	\Phi_{32,\tau} (x,t)&=-x_2 \exp \left( -\frac{\tau^2}{4 C_2 t} \right)  \left( \vphantom{\frac{\left(1-\exp \left( \frac{\tau^2}{4 C_2 t} \right)\right) }{8C_2 \mu t^2 \tau^4 \pi}} \frac{\alpha\mu \tau^4+8C_1(\lambda+2\mu)t\tau^2}{8C_2 \mu t^2 \tau^4 \pi} + \frac{32C_1C_2(\lambda+2\mu)t^2\left(1-\exp \left( \frac{\tau^2}{4 C_2 t} \right)  \right)}{8C_2 \mu t^2 \tau^4 \pi}   \right) , \\
	\Phi_{33,\tau}(x,t) &= \exp \left( -\frac{\tau^2}{4 C_2 t} \right)\frac{8\|x\|^2t-4t\tau^2+c_0 \mu \tau^2(\tau^2-\|x\|^2)}{64 C_2^2 t^4  \pi}.
\end{align} 
\subsection{Theoretical property of an approximate identity}
In this section, we want to show that our source scaling function tensor fulfills the property of an approximate identity. This is the theoretical base for the decorrelation of poroelastic data. In the case of $\Phi_{31,\tau}$, $\Phi_{32,\tau}$ and $\Phi_{33,\tau}$ we have to do a little modification with the Haar scaling function in time to achieve compact support of the source scaling functions also in time. We denote the Haar scaling function by
\begin{align}
	\ph_{t_0} (t) \coloneqq \ph (t/t_0) = \mathcal{X}_{[0,1]} (t/t_0) =\begin{cases}
		1,\quad & 0 \leq t  \leq t_0,\\
		0,\quad & t > t_0
	\end{cases}
\end{align}
and modify the source scaling functions in this way: $\Phi_{\mathrm{new}}(x,t)=\Phi(x,t) \cdot \ph_{t_0}(t)$. We omit the index '$\mathrm{new}$' from now on since we will always use the modified functions. Furthermore, we choose the parameter $t_0$ to be connected with $\tau$ by $t_0=T \cdot\tau$, where $T$ is the end point of our considered time interval.
For the proof of the approximative identity, which is our main theoretical result, we have a look at the volume integrals of the several source scaling functions of the tensor.
\begin{theorem}\label{theorem:source_local_volume}
	We define the volume integral of the source scaling function tensor $\boldsymbol{\Phi}_\tau$ as the tensor
	\begin{align}
		\boldsymbol{V}_{\boldsymbol{\Phi}_\tau}\coloneqq\int_{\R^2} \int_{\R} \boldsymbol{\Phi}_\tau (y,\theta) \ \mathrm{d} \theta \ \dy,
	\end{align}
	that means we have to calculate the volume integral of each component of the tensor $\boldsymbol{\Phi}_\tau$.
	We obtain the following result:
	\begin{align}
		&\left(\boldsymbol{V}_{\boldsymbol{\Phi}_\tau}\right)_{11}=\left(\boldsymbol{V}_{\boldsymbol{\Phi}_\tau}\right)_{22}=1,\ \lim_{\tau \to 0+} \left(\boldsymbol{V}_{\boldsymbol{\Phi}_\tau}\right)_{33}=C_2 c_0\mu,\\
		&\left(\boldsymbol{V}_{\boldsymbol{\Phi}_\tau}\right)_{12}=\left(\boldsymbol{V}_{\boldsymbol{\Phi}_\tau}\right)_{21}=\left(\boldsymbol{V}_{\boldsymbol{\Phi}_\tau}\right)_{13}=\left(\boldsymbol{V}_{\boldsymbol{\Phi}_\tau}\right)_{23}=0,\\
		&  \left(\boldsymbol{V}_{\boldsymbol{\Phi}_\tau}\right)_{31}= \left(\boldsymbol{V}_{\boldsymbol{\Phi}_\tau}\right)_{32}=0.
	\end{align}
	\begin{proof}
		For the proof, we have to distinguish between the three types of functions that we have, namely entries of $\boldsymbol{\Phi}_\tau$ equipped with the delta distribution $\delta_t$, equipped with the derivative of the delta distribution $\delta'_t$ and the spatial and time dependent functions from the last row. Here we will also use the existing symmetries between the source scaling functions and do not calculate each volume integral in detail.
		We start with $\left( \boldsymbol{\Phi}_\tau\right)_{11}$ and use the property from \eqref{eq:conv_deltat}.
		In other words, the time integral in combination with the delta distribution reduces to one and we have only the spatial integral with compact support left.
		Polar coordinates give us
		\begin{align}
			\left(\boldsymbol{V}_{\boldsymbol{\Phi}_\tau}\right)_{11}&=\int_{\B_\tau(0)} -\frac{\lambda+\mu}{\mu} \frac{C_3}{2\pi} \left( -\frac{1}{\tau^2}+C_4 \cdot \left( \frac{6}{\tau^2}-\frac{15x_1^2+5x_2^2}{\tau^4} \right) \right)\notag\\
			&\phantom{=}-\frac{C_3}{2\pi} \left( \frac{-2+4C_4}{\tau^2}+C_4 \cdot \left( -\frac{14x_1^2+2x_2^2}{\tau^4} \right)  \right)+\alpha \frac{C_1}{2\pi} \frac{1}{\tau^2} \left( 2-\frac{3x_1^2+x_2^2}{\tau^2} \right)\ \dx\notag\\
			&=\int_0^\tau \int_0^{2\pi} -\frac{\lambda+\mu}{\mu} \frac{C_3}{2\pi} \left( \frac{6C_4-1}{\tau^2}\cdot r-C_4 \cdot  \frac{15r^2 \cos^2 \ph +5r^2 \sin^2 \ph}{\tau^4}\cdot r  \right)\notag\\
			&\phantom{=}-\frac{C_3}{2\pi} \left( \frac{-2+4C_4}{\tau^2}\cdot r-C_4 \cdot  \frac{14r^2 \cos^2 \ph+2r^2 \sin^2 \ph}{\tau^4}\cdot r   \right)\notag\\
			&\phantom{=}+\alpha \frac{C_1}{2\pi} \frac{1}{\tau^2} \left( 2\cdot r-\frac{3r^2\cos^2 \ph+r^2 \sin^2 \ph}{\tau^2}r \right)\ \mathrm{d} \ph \ \mathrm{d}r.
		\end{align}
	 	Using $\int_0^{2\pi} \cos^2 \ph \ \mathrm{d}\ph=\int_0^{2\pi} \sin^2 \ph \ \mathrm{d} \ph = \pi$, we get
		\begin{align}
			\left(\boldsymbol{V}_{\boldsymbol{\Phi}_\tau}\right)_{11}&=-\frac{\lambda+\mu}{\mu} \frac{C_3}{2\pi} \left( \frac{6C_4-1}{\tau^2} \cdot 2 \pi \cdot \frac{1}{2}\tau^2-C_4\frac{20 \pi}{\tau^4} \cdot\frac{1}{4}\tau^4 \right)\notag\\
			&\phantom{=}-\frac{C_3}{2\pi} \left(  \frac{4C_4-2}{\tau^2} \cdot 2\pi \cdot \frac{1}{2}\tau^2-C_4\cdot \frac{16\pi}{\tau^4}\cdot \frac{1}{4}\tau^4 \right)\notag \\
			&\phantom{=}+\frac{\alpha C_1}{2\pi} \left( \frac{2}{\tau^2}\cdot 2\pi \cdot \frac{1}{2}\tau^2-\frac{4\pi}{\tau^4} \cdot \frac{1}{4}\tau^4 \right)\notag \\
			&=-\frac{\lambda+\mu}{\mu} \cdot \frac{C_3}{2} \left( 6C_4-1-5C_4\right) -\frac{C_3}{2} \left( 4C_4-2-4C_4 \right) +\frac{\alpha C_1}{2}\notag\\
			&=-\frac{\lambda+\mu}{\mu} \cdot \frac{C_3}{2} \left( C_4-1 \right)+C_3+\frac{\alpha C_1}{2}.
		\end{align}
		In a last step, we insert the constants $C_1$, $C_3$ and $C_4$ from \eqref{eq:C1C2} and \eqref{eq:C3C4} and obtain
		\begin{align}
			\left(\boldsymbol{V}_{\boldsymbol{\Phi}_\tau}\right)_{11}&=-\frac{\lambda+\mu}{\mu} \cdot \frac{c_0(\lambda+3\mu)+\alpha^2}{4(c_0(\lambda+2\mu)+\alpha^2)}\cdot \left[ \frac{c_0(\lambda+\mu)+\alpha^2}{c_0(\lambda+3\mu)+\alpha^2} -1 \right]\notag\\
			&\phantom{=}+\frac{c_0(\lambda+3\mu)+\alpha^2}{2(c_0(\lambda+2\mu)+\alpha^2)}+\frac{\alpha^2}{2(c_0(\lambda+2\mu)+\alpha^2)}\notag\\
			&= -\frac{\lambda+\mu}{\mu} \frac{c_0(\lambda+\mu)+\alpha^2-c_0(\lambda+3\mu)-\alpha^2}{4(c_0(\lambda+2\mu)+\alpha^2)}+\frac{c_0(\lambda+3\mu)+2\alpha^2}{2(c_0(\lambda+2\mu)+\alpha^2)}\notag\\
			&=\frac{(-\lambda-\mu)\cdot(-2c_0\mu)+2\mu c_0 (\lambda+3\mu)+4\mu\alpha^2}{4\mu(c_0(\lambda+2\mu)+\alpha^2)}\notag\\
			&=\frac{4\mu c_0 \lambda+8c_0 \mu^2+4\mu \alpha^2}{4\mu(c_0(\lambda+2\mu)+\alpha^2)}\notag\\
			&=1.
		\end{align}
		
		Due to the symmetry of $\Phi_{11,\tau}$ and $\Phi_{22,\tau}$, the same result by doing analogous steps is obtained for the volume integral $\left(\boldsymbol{V}_{\boldsymbol{\Phi}_\tau}\right)_{22}$.
		For $\left({\boldsymbol{\Phi}_\tau}\right)_{12}$, we apply the familiar argument for the delta distribution from above and get for the integral over the spatial part
		\begin{align}
			\int\limits_{\bb} \Phi_{12,\tau} (x) \ \dx &= \int\limits_{\B_\tau(0)} \Phi_{12,\tau}(x)\  \dx\notag \\
			&=\int\limits_{\B_\tau(0)}\frac{x_1 x_2}{\tau^4 \pi}\left( \frac{\lambda+\mu}{\mu} \cdot 5 C_3 C_4  +6C_3 C_4-\alpha C_1 \right)\ \dx \notag\\
			&=\frac{1}{\tau^4 \pi} \left( \frac{\lambda+\mu}{\mu} \cdot 5 C_3 C_4  +6C_3 C_4-\alpha C_1 \right)\int\limits_0^{2\pi}\int\limits_0^\tau r^3 \sin(\ph) \cos(\ph) \ \mathrm{d}r \ \mathrm{d} \ph \notag\\
			&=0.
		\end{align}
		With the help of symmetry arguments, we also have $\left(\boldsymbol{V}_{\boldsymbol{\Phi}_\tau}\right)_{21}=0$.
		For the components $\left({\boldsymbol{\Phi}_\tau}\right)_{13}$ and $\left({\boldsymbol{\Phi}_\tau}\right)_{23}$, we have to split two cases (depending on the shape of the delta distribution that means with or without derivative).
		In the case of $\delta_t$, we can apply the same argument as above and apply polar coordinates to the remaining integral over the spatial part
		\begin{align}
			\int_{\B_\tau (0)} \frac{4x_1 C_1}{\tau^3\pi} =\frac{4C_1}{\tau^3 \pi}\int_0^\tau \int_0^{2\pi} r^2 \cos \ph \ \mathrm{d}\ph \ \mathrm{d}r =0. \notag
		\end{align}
		For the second part with $\delta_t'$, we get directly from \eqref{eq:conv_deltastricht} that this integral is zero.
		Due to the similarity of $\left( \boldsymbol{\Phi}_\tau \right)_{23}$ and $\left( \boldsymbol{\Phi}_\tau \right)_{13}$, we get with analogous arguments that the volume integral of $\left( \boldsymbol{\Phi}_\tau \right)_{23}$ vanishes as well.\\
		Eventually, the last row of $\boldsymbol{\Phi}_\tau$ with the time-dependent functions is left.
		In the case of the function $\left(\boldsymbol{\Phi}_\tau\right)_{31}$ and its some symmetric properties based on the factor $x_1$, we can separate the spatial integral from the time integral.
		\begin{align}
			\int\limits_{\mathcal{B}} \int\limits_0^{t_0} \Phi_{31,\tau} \ \mathrm{d} t\ \mathrm{d} x &=-\int_{\mathbb{B}_\tau(0)} x_1 \ \mathrm{d} x
			 \int_0^{t_0}\exp \left( -\frac{\tau^2}{4 C_2 t} \right) \\ \notag
			 &\qquad \times \left( \vphantom{\frac{\left(1-\exp \left( \frac{\tau^2}{4 C_2 t} \right)\right) }{8C_2 \mu t^2 \tau^4 \pi}} \frac{\alpha \mu \tau^4+8C_1(\lambda+2\mu)t\tau^2  }{8 C_2 \mu t^2 \tau^4 \pi}
			 + \frac{ 32C_1 C_2 (\lambda+2\mu) t^2\left(1-\exp \left( \frac{\tau^2}{4 C_2 t} \right)\right)  }{8 C_2 \mu t^2 \tau^4 \pi}\right) \ \mathrm{d}t \\ \notag
			&=-\int_0^\tau \int_0^{2\pi} r \cos (\ph) r \ \mathrm{d}\ph \ \mathrm{d} r
			\int_0^{t_0}\exp \left( -\frac{\tau^2}{4 C_2 t} \right) \\ \notag
			&\qquad \times \left( \vphantom{\frac{\left(1-\exp \left( \frac{\tau^2}{4 C_2 t} \right)\right) }{8C_2 \mu t^2 \tau^4 \pi}} \frac{\alpha \mu \tau^4+8C_1(\lambda+2\mu)t\tau^2  }{8 C_2 \mu t^2 \tau^4 \pi}
			 + \frac{ 32C_1 C_2 (\lambda+2\mu) t^2\left(1-\exp \left( \frac{\tau^2}{4 C_2 t} \right)\right)  }{8 C_2 \mu t^2 \tau^4 \pi}\right) \ \mathrm{d}t \notag \\
			&=0 \cdot \int_0^{t_0}\exp \left( -\frac{\tau^2}{4 C_2 t} \right)\\ \notag
			&\qquad \times \left( \vphantom{\frac{\left(1-\exp \left( \frac{\tau^2}{4 C_2 t} \right)\right) }{8C_2 \mu t^2 \tau^4 \pi}} \frac{\alpha \mu \tau^4+8C_1(\lambda+2\mu)t\tau^2  }{8 C_2 \mu t^2 \tau^4 \pi}
		 + \frac{ 32C_1 C_2 (\lambda+2\mu) t^2\left(1-\exp \left( \frac{\tau^2}{4 C_2 t} \right)\right)  }{8 C_2 \mu t^2 \tau^4 \pi}\right) \ \mathrm{d}t \notag\\
			&= 0.
		\end{align}	
		With this it also follows that $\left(\boldsymbol{V}_{\boldsymbol{\Phi}_\tau}\right)_{32}=0$ due to the symmetry. For the last function, we need the following integrals
		\begin{align}
			&\int_{\mathbb{B}_\tau(0)} \|x\|^2 \ \mathrm{d}x = \int_0^{2\pi} \int_0^\tau r^2 \cdot r \ \mathrm{d}r\ \mathrm{d} \ph = 2\pi\cdot \frac{1}{4} r^4\big|_0^\tau = \frac{\pi}{2} \tau^4, \\
			&\int_{\mathbb{B}_\tau(0)} 1 \ \mathrm{d}x = \pi \tau^2.
		\end{align}
		With this and \cref{remark:exp_integral}, we can calculate the integral over $\left({\boldsymbol{\Phi}_\tau}\right)_{33}$ by
		\begin{align}
			\int_{\mathcal{B}} \int_0^{t_0} &\Phi_{33,\tau} \ \mathrm{d}t\ \mathrm{d}x \notag \\
			&= \int_0^{t_0}\int_{\B_\tau (0)}\exp \left( -\frac{\tau^2}{4 C_2 t} \right) \frac{ 4t(2\|x\|^2-\tau^2)+c_0 \mu \tau^2 (\tau^2-\|x\|^2)}{64 C_2^2 t^4 \pi} \ \mathrm{d}x\ \mathrm{d}t\notag\\
			&= \int_0^{t_0}\exp \left( -\frac{\tau^2}{4 C_2 t} \right) \frac{ 4t \left(  \tau^4 \pi-\pi \tau^4 \right)+c_0 \mu \tau^2 (\pi \tau^4-\frac{1}{2}\tau^4\pi)}{64 C_2^2 t^4 \pi}  \ \mathrm{d}t\notag\\
			&=\frac{1}{64C_2^2}\frac{c_0 \mu \tau^6}{2} \int_0^{t_0}\exp \left( -\frac{\tau^2}{4 C_2 t} \right) \frac{1}{t^4}  \ \mathrm{d}t\notag\\
			&=\frac{1}{64C_2^2}\frac{c_0 \mu \tau^6}{2} \cdot \frac{4C_2 \exp \left( -\frac{\tau^2}{4 C_2 t_0} \right)\left(32C_2^2t_0^2+8C_2t_0\tau^2+\tau^4\right)}{t_0^2 \tau^6}.
		\end{align}
		In a last step, we have to take into account the coupling between $\tau$ and $t_0$.
		For a constant $T>0$ (which is the length of our considered time interval) and $t_0=T \tau$, we get
		\begin{align}
			\lim_{\tau \to 0+} V_{\Phi_{33,\tau}} = \frac{1}{64C_2^2}\frac{c_0 \mu }{2} \cdot 4C_2\cdot 32C_2^2  = C_2 c_0 \mu. \notag \tag*{\qedhere} 
		\end{align}
	\end{proof}
\end{theorem}
For a better comparison of the convolution results in practice, it would be better to modify $\Phi_{33,\tau}$ in such a way that the volume integral is $1$. For this reason, we define the new function $\Phi_{33,\tau}^{\new}$ by
\begin{align}
	\Phi_{33,\tau}^{\new}=\frac{1}{\left(\boldsymbol{V}_{\boldsymbol{\Phi}_\tau}\right)_{33}} \Phi_{33,\tau}. \label{eq:volume_phi33}
\end{align}

\noindent
This is necessary on the one hand to prove \cref{theorem:identity} and on the other hand for a better comparison of the decorrelation for different parameters $\tau$.
In the further considerations we always use the newly defined function although we omit the upper index '$\mathrm{new}$' for a better readability.

Before we are able to show the theorem of an approximate identity, we are left to prove another lemma, which we need for the main theorem.
\begin{lemma} \label{lemma:phi11_pos}
	The positive part of $\Phi_{11,\tau}$ achieves its maximum at $(\tau,0)$ or $(-\tau,0)$.
	\begin{proof}
		We start by a rearrangement of $\Phi_{11,\tau}$ to the following form
		\begin{align}
			\Phi_{11,\tau}&=\left( -\frac{\lambda+\mu}{\mu} \frac{C_3}{2\pi} \frac{1}{\tau^2} (6C_4-1)-\frac{C_3}{2\pi} \frac{1}{\tau^2} (4C_4-2)+\frac{\alpha C_1}{2 \pi}\frac{2}{\tau^2}\right) \notag\\
			&\phantom{=}+ x_1^2 \left( \frac{\lambda+\mu}{\mu} \frac{C_3 \cdot C_4}{2 \pi \tau^4}\cdot 15+\frac{C_3\cdot C_4}{2 \pi \tau^4}\cdot 14-\frac{\alpha C_1}{2\pi \tau^4}\cdot 3 \right)\notag\\
			&\phantom{=}+  x_2^2 \left( \frac{\lambda+\mu}{\mu} \frac{C_3 \cdot C_4}{2 \pi \tau^4}\cdot 5+\frac{C_3\cdot C_4}{2 \pi \tau^4}\cdot 2-\frac{\alpha C_1}{2\pi \tau^4} \right)\notag\\
			&\eqqcolon D_0 + x_1^2 D_1+x_2^2 D_2.
		\end{align}
		We introduce $D_0\eqqcolon \frac{1}{\tau^2} d_0$, $D_1\eqqcolon \frac{1}{\tau^4} d_1$ and $D_2\eqqcolon \frac{1}{\tau^4} d_2$ and show that $d_0<0$ and $d_1,\ d_2 >0$. For this purpose, we insert the constants from \eqref{eq:C1C2} and \eqref{eq:C3C4}:
		\begin{align}
			d_0&=-\frac{\lambda+\mu}{\mu} \frac{C_3}{2\pi}  (6C_4-1)-\frac{C_3}{2\pi}  (4C_4-2)+\frac{\alpha C_1}{ \pi}\notag\\
			&= \frac{C_3 C_4}{2\pi} \left( -\frac{\lambda+\mu}{\mu} \cdot 6 -4 \right) +\frac{C_3}{2\pi}\left( \frac{\lambda+\mu}{\mu}+2\right) +\frac{\alpha C_1}{\pi}\notag\\
			&=\frac{1}{2\pi} \frac{c_0(\lambda+\mu)+\alpha^2}{2(c_0(\lambda+2\mu)+\alpha^2)}\cdot \frac{-6\lambda-10\mu}{\mu}+\frac{1}{2\pi} \frac{c_0(\lambda+3\mu)+\alpha^2}{2(c_0(\lambda+2\mu)+\alpha^2)}\cdot \frac{\lambda+3\mu}{\mu} \notag \\
			&\phantom{=}+\frac{1}{\pi} \frac{\alpha^2}{c_0(\lambda+2\mu)+\alpha^2}\notag\\
			&= \frac{1}{2\pi}\frac{(c_0(\lambda+\mu)+\alpha^2)(-6\lambda-10\mu)+(c_0(\lambda+3\mu)+\alpha^2)(\lambda+3\mu)+4\alpha^2\mu}{2\mu (c_0(\lambda+2\mu)+\alpha^2)} \notag\\
			&=\frac{1}{2\pi}\frac{-5c_0\lambda^2-10c_0 \lambda\mu-c_0\mu^2-5\lambda\alpha^2-3\mu\alpha^2}{2\mu (c_0(\lambda+2\mu)+\alpha^2)} <0.
		\end{align}
		In analogy to the previous considerations, we obtain for $d_1$ and $d_2$
		\begin{align}
			d_1&=\frac{1}{2\pi} \left(  \frac{15\lambda+29\mu}{\mu}\cdot C_3 C_4-3\alpha C_1 \right) \notag\\
			&= \frac{1}{2\pi} \frac{(c_0(\lambda+\mu)+\alpha^2)\cdot 15 \lambda+29\mu c_0(\lambda+\mu)+23\mu \alpha^2}{2\mu (c_0(\lambda+2\mu)+\alpha^2)} >0,\\
			d_2&=\frac{1}{2\pi} \left( \frac{5\lambda+7\mu}{\mu} \cdot C_3 C_4-\alpha C_1\right)\notag\\
			&= \frac{1}{2\pi} \frac{(c_0(\lambda+\mu)+\alpha^2)\cdot 5\lambda +c_0(\lambda+\mu) \cdot 7 \mu +5 \mu \alpha^2}{2\mu (c_0(\lambda+2\mu)+\alpha^2)} >0.
		\end{align}
		This result also reveals that $d_1>d_2$.
		That means we have an optimization problem where we want to find the maximum of the function $\Phi_{11,\tau}(x)=D_0+x_1^2D_1+x_2^2D_2$ under the given constraint $g(x)\coloneqq x_1^2+x_2^2 -\tau^2 \leq 0$. This is done by determining the Karush-Kuhn-Tucker (KKT) points. This results in minimizing $F(x)=-\Phi_{11,\tau}$ under the constraint above.
		We get the following conditions
		\begin{align}
			\nabla F(x)+ u \nabla g(x)&\overset{!}{=}0 \\
			\Leftrightarrow\  -2x_1 D_1+u 2x_1 &= 0 \label{eq:kkt1}\\
			-2x_2 D_2+u 2x_2 &=0\label{eq:kkt2}\\
			\text{and}\phantom{ug(x)}  u\left( x_1^2+x_2^2-\tau^2\right)&=0\label{eq:kkt3}\\
			\text{and} \phantom{u(x_1^2+x_2^2-\tau^2)g(x)}u& \geq 0\label{eq:kkt4} \\
			\text{and} \phantom{u(x_1^2+x_2^2-\tau^2)u} g(x) &\leq 0\label{eq:kkt5}.
		\end{align}
		We get from \eqref{eq:kkt1} that $x_1=0 $ or $ u=D_1$ and from \eqref{eq:kkt2} that $x_2=0$ or $u=D_2$. Combining these conditions with \eqref{eq:kkt3}-\eqref{eq:kkt5}, we arrive at the following cases
		\begin{enumerate}[a)]
			\item \underline{$x_1=0$ and $x_2=0$:} With \eqref{eq:kkt3}, we get $u=0$ and \eqref{eq:kkt4} and \eqref{eq:kkt5} are fulfilled. We obtain the point $(0,0,0)$.
			\item \underline{$x_1=0$ and $u=D_2$:} With \eqref{eq:kkt3} we find $u(x_2^2-\tau^2)=0$ and therefore $u=0$ (contradiction to $u=D_2$) or $x_2=\pm \tau$. \eqref{eq:kkt4} and \eqref{eq:kkt5} are fulfilled. We obtain the points $(0,\pm \tau,D_2)$.
			\item \underline{$x_2=0$ and $u=D_1$:} With \eqref{eq:kkt3}, we get $u(x_1^2-\tau^2)=0$ and therefore $u=0$ (contradiction to $u=D_1$) or $x_1=\pm \tau$. \eqref{eq:kkt4} and \eqref{eq:kkt5} are fulfilled and we obtain the points $(\pm \tau,0,D_1)$.
		\end{enumerate}
		Inserting these KKT-points in the functions $F$ and $\Phi_{11,\tau}$, we get
		\begin{align}
			&F(0,0)=-D_0,\quad F(0,\pm\tau)=-D_0-\tau^2D_2,\quad F(\pm\tau,0)=-D_0-\tau^2 D_1, \\
			&\Phi_{11,\tau}(0,0)=D_0,\quad \Phi_{11,\tau}(0,\pm \tau)=D_0+\tau^2D_2,\quad \Phi_{11,\tau}(\pm\tau,0)=D_0+\tau^2D_1.
		\end{align}
		Since we have $D_1>D_2>0$, the function has to achieve its maximum at the points $(\tau,0)$ and $(-\tau,0)$.
		To be exact: The side condition describes a compact disc and the function $\Phi_{11,\tau}$ is continuous, which results in the existence of a minimum and a maximum. Therefore, we have a maximum in one of the KKT-points.
	\end{proof}
\end{lemma}
\noindent
After this preparatory work, we can finally present our central result for our calculated source scaling function tensor $\boldsymbol{\Phi}_\tau$ from \eqref{eq:source_functions_tensor}, namely the approximate identity.
\begin{theorem}\label{theorem:identity}
	Let $\mathcal{B}$ be a regular region in $\R^2$ and $f:\overline{\bb}\times \R \to \R^3$ be continuously differentiable. Then
	\begin{align}
		\lim_{\substack{\tau \to 0 \\ \tau >0}} \int\limits_{\mathcal{B}} \int\limits_{\R} \boldsymbol{\Phi}_{\tau} (x-y,t-\theta)f(y,\theta)\ \mathrm{d} y \ \mathrm{d} \theta = f(x,t) \label{eq:centralresult}
	\end{align}
	is fulfilled for all $x \in \mathcal{B}$, $t >0$ and our constructed $\boldsymbol{\Phi}_{\tau}$ is a scaling function providing an approximate identity.
\end{theorem}
\begin{proof}
	A more detailed notation of the approximate identity \eqref{eq:centralresult} gives us
	\begin{align}
		\lim_{\substack{\tau \to 0 \\ \tau >0}} \int\limits_{\mathcal{B}} \int\limits_{\R} &\Phi_{11,\tau}(x-y) \delta_t f_1(y,\theta)+\Phi_{12,\tau} (x-y) \delta_t f_2(y,\theta)\notag \\
		&+\left( \Phi_{13,\tau}^{1}(x-y) \delta_t'+\Phi_{13,\tau}^2(x-y) \delta_t \right) f_3(y,\theta)\ \mathrm{d} \theta\ \mathrm{d}y  = f_1(x,t) \label {eq:decor1}, \\
		\lim_{\substack{\tau \to 0 \\ \tau >0}} \int\limits_{\mathcal{B}} \int\limits_{\R} &\Phi_{21,\tau}(x-y) \delta_t f_1(y,\theta)+\Phi_{22,\tau} (x-y) \delta_t f_2(y,\theta) \notag\\
		&+\left( \Phi_{23,\tau}^1(x-y) \delta'_t+\Phi_{23,\tau}^2(x-y) \delta_t \right) f_3(y,\theta) \ \mathrm{d} \theta\ \mathrm{d}y = f_2(x,t) \label{eq:decor2}, \\
		\lim_{\substack{\tau \to 0 \\ \tau >0}} \int\limits_{\mathcal{B}} \int\limits_{\R} &\Phi_{31,\tau}(x-y,t-\theta)  f_1(y,\theta)+\Phi_{32,\tau} (x-y,t-\theta)  f_2(y,\theta) \notag\\
		&+\Phi_{33,\tau}(x-y,t-\theta)f_3(y,\theta)\ \mathrm{d} \theta\ \mathrm{d}y  = f_3(x,t)\label{eq:decor3}.
	\end{align}
	
	Again using \eqref{eq:conv_deltat} and \eqref{eq:conv_deltastricht}, we can change the equations above to the following
	\begin{align}
		\lim_{\substack{\tau \to 0 \\ \tau >0}} \int\limits_{\mathcal{B}}  &\Phi_{11,\tau}(x-y)  f_1(y,t)+\Phi_{12,\tau} (x-y)  f_2(y,t)\notag \\
		&- \Phi_{13,\tau}^{1}(x-y) \frac{\partial}{\partial t}f_3(y,t)+\Phi_{13,\tau}^2(x-y)   f_3(y,t)\ \mathrm{d}y = f_1(x,t) \label {eq:decor1b}, \\
		\lim_{\substack{\tau \to 0 \\ \tau >0}} \int\limits_{\mathcal{B}} &\Phi_{21,\tau}(x-y) f_1(y,t)+\Phi_{22,\tau} (x-y) f_2(y,t) \notag\\
		&- \Phi_{23,\tau}^1(x-y) \frac{\partial}{\partial t}f_3(y,t)+\Phi_{23,\tau}^2(x-y)   f_3(y,t)\ \mathrm{d}y= f_2(x,t) \label{eq:decor2b}, \\
		\lim_{\substack{\tau \to 0 \\ \tau >0}} \int\limits_{\mathcal{B}} \int\limits_{\R} &\Phi_{31,\tau}(x-y,t-\theta)  f_1(y,\theta)+\Phi_{32,\tau} (x-y,t-\theta)  f_2(y,\theta) \notag\\
		&+\Phi_{33,\tau}(x-y,t-\theta)f_3(y,\theta)\ \mathrm{d}y \ \mathrm{d} \theta = f_3(x,t)\label{eq:decor3b}.
	\end{align}
	Our starting point are the several parts of \eqref{eq:decor1} and we are guided by the technique given in \cite{blick_eberle}.
	We have $x \in \bb$ and $\bb$ is open. Hence, there exists a $\tau_0$ such that $\bb \cap \B_\tau (x)=\B_\tau (x)$ for all $0 <\tau \leq \tau_0$.
	We now write the equation for the approximate identity with time dependence by utilizing the compact support of the functions as
	\begin{align}
		\int_{\bb} \int_{\R} \boldsymbol{\Phi}_\tau (x-y,t-\theta)  f(y,\theta) \ \mathrm{d} \theta \ \dy
		&= \int_{\B_{\tau}(x)} \int_{t-t_0}^{t} \boldsymbol{\Phi}_\tau(x-y,t-\theta)f(y,\theta )\ \mathrm{d} \theta \ \dy \notag\\
		&= \left(  \sum_{j=1}^3 \int_{\B_\tau(x)} \int_{t-t_0}^{t} \left( \boldsymbol{\Phi}_\tau \right)_{ij} (x-y,t-\theta)f_j(y,\theta) \ \mathrm{d} \theta \ \dy \right)_{i=1,2,3} . \label{eq:approx_id_sum}
	\end{align}
	Splitting $(\boldsymbol{\Phi}_\tau)_{ij}$ into its positive and negative parts results in
	\begin{align}
		(\boldsymbol{\Phi}_\tau)_{ij}^+(x,t)=\begin{cases}
			(\boldsymbol{\Phi}_\tau)_{ij}(x,t),\quad & (\boldsymbol{\Phi}_\tau)_{ij}(x,t)\geq 0\\
			0, \quad & (\boldsymbol{\Phi}_\tau)_{ij}(x,t)<0
		\end{cases},\\
		(\boldsymbol{\Phi}_\tau)^-_{ij}(x,t)=\begin{cases}
			(\boldsymbol{\Phi}_\tau)_{ij}(x,t),\quad & (\boldsymbol{\Phi}_\tau)_{ij}(x,t) \leq 0\\
			0,\quad & (\boldsymbol{\Phi}_\tau)_{ij}(x,t)>0
		\end{cases}.
	\end{align}
	With this we have
	\begin{align}
		\int_{\B_\tau(x)} \int_{t-t_0}^{t} \left(\boldsymbol{\Phi}_\tau \right)_{ij} (x-y,t-\theta) f_j(y,\theta)\ \mathrm{d} \theta \ \dy
		&= \int_{\B_\tau(x)} \int_{t-t_0}^{t} \left( \boldsymbol{\Phi}_\tau \right)_{ij}^+ (x-y,t-\theta) f_j(y,\theta)\ \mathrm{d} \theta \ \dy \notag\\
		&\phantom{=}+ \int_{\B_\tau(x)} \int_{t-t_0}^{t} \left( \boldsymbol{\Phi}_\tau \right)_{ij}^- (x-y,t-\theta) f_j(y,\theta)\ \mathrm{d} \theta \ \dy.
	\end{align}
	We make use of the continuity of $f$ and the integrability of $\left(\boldsymbol{\Phi}_\tau \right)_{ij}^+$ and $\left(\boldsymbol{\Phi}_\tau \right)_{ij}^-$. Together with the fact that the positive and negative part do not change their sign in $\B_\tau(x) \times [t-t_0,t]$, we can apply the mean value theorem of integration.
	We have to distinguish two cases again, the parts of $\Phi_\tau$ with $\delta_t$ or $\delta_t'$ and the spatially and temporally dependent functions. Starting with the first case and applying the mean value theorem only for the spatial part while considering \eqref{eq:decor1b} and \eqref{eq:decor2b}, we obtain
	\begin{align}
		\int_{\B_\tau(x)} \Phi_{ij,\tau} (x-y) f_j(y,t) \ \dy=f_j(\xi_1,t) \int_{\B_\tau (x)} \Phi_{ij,\tau}^+ (x-y)\ \dy
		+f_j(\xi_2,t) \int_{\B_{\tau}(x)} \Phi_{ij,\tau}^- (x-y)\ \dy,\ i=1,2,\ j=1,2,3. \label{eq:mws_poro_local}
	\end{align}
	The mean value theorem guarantees the existence of such $\xi_1,\xi_2 \in \B_\tau(x)$.
	With the help of \cref{theorem:source_local_volume}, we get 
	\begin{align}
		\int_{\B_\tau(x)}   \Phi_{ij,\tau}^+ (x-y) \  \dy + \int_{\B_\tau(x)} \Phi_{ij,\tau}^- (x-y) \  \dy=\delta_{ij},\ i=1,2, j=1,2,3.
	\end{align}
	Rearranging this equation to
	\begin{align}
		\int_{\B_\tau(x)} \Phi_{ij,\tau}^- (x-y)  \ \dy=\delta_{ij}-\int_{\B_\tau(x)}   \Phi_{ij,\tau}^+ (x-y)  \ \dy
	\end{align}
	and substituting it into \eqref{eq:mws_poro_local}, we get
	\begin{align}
		\int_{\B_\tau(x)}  &  \Phi_{ij,\tau} (x-y) f_j(y,t) \ \dy =f_j(\xi_2,t)\delta_{ij}+\left( f_j(\xi_1,t)-f_j(\xi_2,t) \right)  \int_{\B_\tau(x)}    \Phi_{ij,\tau}^+ (x-y) \ \dy. \label{eq:maintheorem}
	\end{align}
	Now we have to estimate
	\begin{align}
		\int_{\B_\tau(x)}   \Phi_{ij,\tau}^+ (x-y,t-\theta) \ \dy \leq C
	\end{align}
	for a positive constant $C$ independent of $\tau$. Once we achieve this, we will be able to derive
	\begin{align}
		\lim_{\substack{\tau \to 0 \\ \tau >0}} \int_{\B_\tau (x)}   \Phi_{ij,\tau} (x-y) f_j(y,t) \ \dy=\delta_{ij}f_j(x,t)
	\end{align}
	since $\xi_1,\xi_2 \in \B_\tau (x)$ and $f_j$ is continuous.
	This will prove our theorem for the functions with $\delta_t$-part, because then the difference $(f_j(\xi_1,t)-f_j(\xi_2,t))$ shrinks to 0 for $\tau \to 0+$.
	This can be done in analogy for the parts with $\delta_t'$ but in this case we have the derivative with respect to $t$ for $f_j$ instead of $f_j$.
	The same technique can be applied for the spatially and temporally dependent functions from the last row of $\boldsymbol{\Phi}_\tau$, which we will also depict here.
	For this case we do not have to go back to the beginning of the proof since the latter was worded in general terms. We proceed at the point where we applied the mean value theorem, which we now apply for the spatial and time integral.
	
	This guarantees the existence of $(\xi_1,\eta_1),\ (\xi_2,\eta_2) \in \B_\tau(x) \times [0,t_0]$, such that
	\begin{align}
		\int_{\B_\tau(x)} \int_{t-t_0}^{t} \left( \boldsymbol{\Phi}_\tau \right)_{ij} (x-y,t-\theta)  f_j(y,\theta) \ \mathrm{d} \theta \  \dy
		&= f_j(\xi_1,\eta_1) \int_{\B_\tau(x)} \int_{t-t_0}^{t} \left( \boldsymbol{\Phi}_\tau \right)_{ij}^+ (x-y,t-\theta)\ \mathrm{d} \theta \  \dy\notag\\
		&\phantom{=}+f_j(\xi_2,\eta_2) \int_{\B_\tau(x)} \int_{t-t_0}^{t} \left( \boldsymbol{\Phi}_\tau \right)_{ij}^- (x-y,t-\theta)\ \mathrm{d} \theta \  \dy. \label{eq:mws_poro}
	\end{align}
 Together with \cref{theorem:source_local_volume} and \eqref{eq:volume_phi33}, we get
	\begin{align}
		\int_{\B_\tau(x)}  \int_{t-t_0}^{t} \left( \boldsymbol{\Phi}_\tau \right)_{ij}^+ (x-y,t-\theta) \ \mathrm{d} \theta \  \dy + \int_{\B_\tau(x)}  \int_{t-t_0}^{t}\left( \boldsymbol{\Phi}_\tau \right)_{ij}^- (x-y,t-\theta)\ \mathrm{d}\theta \  \dy=\delta_{ij}.
	\end{align}
	We can rearrange this equation to
	\begin{align}
		\int_{\B_\tau(x)}  \int_{t-t_0}^{t} \left( \boldsymbol{\Phi}_\tau \right)_{ij}^- (x-y,t-\theta) \ \mathrm{d} \theta \ \dy=\delta_{ij}-\int_{\B_\tau(x)}  \int_{t-t_0}^{t} \left( \boldsymbol{\Phi}_\tau \right)_{ij}^+ (x-y,t-\theta) \ \mathrm{d} \theta \ \dy.
	\end{align}
	Now substituting this equation into \eqref{eq:mws_poro}, we have
	\begin{align}
		\int_{\B_\tau(x)}&  \int_{t-t_0}^{t}\left( \boldsymbol{\Phi}_\tau \right)_{ij} (x-y,t-\theta) f_j(y,\theta)\ \mathrm{d} \theta \ \dy \notag\\
		&=f_j(\xi_2,\eta_2)\delta_{ij}+\left( f_j(\xi_1,\eta_1)-f_j(\xi_2,\eta_2) \right)  \int_{\B_\tau(x)}  \int_{t-t_0}^{t} \left( \boldsymbol{\Phi}_\tau \right)_{ij}^+ (x-y,t-\theta)\ \mathrm{d} \theta \ \dy.
	\end{align}
	In a last step, we have to estimate again
	\begin{align}
		\int_{\B_\tau(x)}  \int_{t-t_0}^{t} \left( \boldsymbol{\Phi}_\tau \right)_{ij}^+ (x-y,t-\theta)\ \mathrm{d} \theta \ \dy \leq C
	\end{align}
	for a positive constant $C$ independent of $\tau$. It follows that
	\begin{align}
		\lim_{\substack{\tau \to 0 \\ \tau >0}} \int_{\B_\tau (x)} \int_{t-t_0}^{t} \left( \boldsymbol{\Phi}_\tau \right)_{ij} (x-y,t-\theta) f_j(y,\theta)\ \mathrm{d} \theta \ \dy=\delta_{ij}f_j(x,t)
	\end{align}
	since $(\xi_1,\eta_1)$,$(\xi_2,\eta_2) \in \B_\tau (x) \times [t-t_0,t]$, $f_j$ is continuous and we have to take into account the coupling between $\tau$ and $t_0$ that means $t_0$ goes to zero in the same manner as $\tau$. Summarizing all results, we get
	\begin{align}
		\lim_{\substack{\tau \to 0 \\ \tau >0}} \int_{\bb} \int_\R \boldsymbol{\Phi}_\tau (x-y,t-\theta )f(y,\theta)\ \mathrm{d} \theta \ \dy=f(x,t).
	\end{align}
	So the last point to complete the proof is to show that the integral over the positive part of each source scaling function is restricted by a positive constant $C$ which is independent of $\tau$. Therefore we have again a look at each function separately.
	Let us start once more with the source scaling functions equipped with $\delta_t$ and $\delta_t'$ and use \eqref{eq:conv_deltat} and \eqref{eq:conv_deltastricht}. We start with $\Phi_{11,\tau}$ and estimate
	\begin{align}
		\int_{\B_\tau (x)} \left( \Phi_{11,\tau} \right)^+ (x-y) \ \dy \leq \max_{y \in \mathbb{B}_\tau (x)} \left| \left( \Phi_{11,\tau} \right)^+ (x-y) \right| \underbrace{\int_{\B_\tau (x)} 1\  \dy}_{=\pi \tau^2}.
	\end{align}
	We know from \cref{lemma:phi11_pos} that $\Phi_{11,\tau}$ achieves its maximum at $(\pm\tau,0)$. Inserting this leads us to
	\begin{align}
		\Phi_{11,\tau}(\pm\tau,0)&=-\frac{\lambda+\mu}{\mu} \frac{C_3}{2\pi} \left( \frac{6C_4-1}{\tau^2}-C_4\frac{15\tau^2}{\tau^4}\right) \notag\\
		&\phantom{=}-\frac{C_3}{2\pi} \left( \frac{4C_4-2}{\tau^2}-C_4 \frac{14\tau^2}{\tau^4} \right) +\frac{\alpha C_1}{2\pi} \left( \frac{2}{\tau^2}-\frac{3\tau^2}{\tau^4} \right) \notag \\
		&=  \mathcal{O} \left(\tau^{-2}\right) \text{ as } \tau \to 0+.
	\end{align}
	Hence, we can estimate the integral over the positive part by a constant. We continue with
	\begin{align}
		\int_{\B_\tau (x)} \left( \Phi_{12,\tau} \right)^+ (x-y) \ \dy=\int_{\B_\tau (0)} \left( \Phi_{12,\tau} \right)^+ (y) \ \dy
	\end{align}
	and obtain for the corresponding constants
	\begin{align}
		\frac{\lambda+\mu}{\mu} \frac{C_3C_4}{\pi} \cdot 5+\frac{C_3 C_4}{\pi} \cdot 6 -\alpha\frac{C_1}{\pi}
		&= \frac{C_3 C_4}{\pi} \left(5 \frac{\lambda+\mu}{\mu}+6 \right) -\frac{\alpha C_1}{\pi} \notag\\
		&= \frac{1}{\pi} \left( C_3 C_4 \frac{5\lambda+11\mu}{\mu} -\alpha C_1\right)\notag\\
		&= \frac{1}{\pi} \left( \frac{c_0(\lambda+\mu)+\alpha^2}{2(c_0(\lambda+2\mu)+\alpha^2)} \cdot \frac{5\lambda+11\mu}{\mu}-\frac{\alpha^2}{c_0(\lambda+2\mu)+\alpha^2} \right)\notag\\ &>0.
	\end{align}
	Since the combination of constants above is positive and the function possesses a symmetry, we can have a look at $\Phi_{12,\tau}$ for $y_1,y_2 >0$. The use of polar coordinates gives us
	\begin{align}
		\int_{\substack{\B_\tau (0)\\ y_1,y_2>0}}& \left( \Phi_{12,\tau} \right)^+ (y) \ \dy \notag \\
		&=\left( 5\frac{\lambda+\mu}{\mu} \frac{C_3 C_4}{\pi \tau^4} +6 \frac{C_3C_4}{\pi \tau^4} -\frac{\alpha C_1}{\pi \tau^4} \right) \int_0^{\pi/2} \int_0^\tau \left( r^2 \sin \ph \cos \ph \right)\cdot r \ \mathrm{d}r\ \mathrm{d}\ph \notag \\
		&=\left( 5\frac{\lambda+\mu}{\mu} \frac{C_3 C_4}{\pi \tau^4} +6 \frac{C_3C_4}{\pi \tau^4} -\frac{\alpha C_1}{\pi \tau^4} \right) \cdot \frac{1}{2} \cdot \frac{1}{4} \tau^4.
	\end{align}
	This result implies that the positive part can also be estimated by a constant.
	Now we continue with the two components of $\Phi_{13,\tau}$. We are able to show that the function $\Phi_{13,\tau}^1$ without the factor $x_1$ is positive. For this purpose, we utilize the fact that $\|x\|<\tau$:
	\begin{align}
		&\frac{c_0 \mu C_1}{2\pi}\frac{1}{\tau^2} \left( 2-\frac{\|x\|^2}{\tau^2} \right)+\frac{\alpha C_3}{2\pi} \frac{1}{\tau^2} \left( 6C_4-1-5C_4 \frac{\|x\|^2}{\tau^2} \right)\notag\\
		&>\frac{c_0\mu C_1}{2\pi} \frac{1}{\tau^2}  \left( 2-\frac{\tau^2}{\tau^2} \right) +\frac{\alpha C_3}{2\pi} \frac{1}{\tau^2} \left(  6C_4-1-5C_4 \frac{\tau^2}{\tau^2} \right)\notag\\
		&=\frac{1}{2\pi\tau^2} \left( c_0\mu C_1+\alpha C_3 (C_4-1)\right)\notag\\
		&= \frac{1}{2\pi\tau^2}\left( c_0 \mu \frac{\alpha}{c_0(\lambda+2\mu)+\alpha^2}+\alpha \frac{c_0(\lambda+3\mu)+\alpha^2}{2(c_0(\lambda+2\mu)+\alpha^2)} \cdot \left( \frac{c_0(\lambda+\mu)+\alpha^2}{c_0(\lambda+3\mu)+\alpha^2}-1 \right)\right)\notag\\
		&=\frac{1}{2\pi\tau^2} \left( c_0 \mu \frac{\alpha}{c_0(\lambda+2\mu)+\alpha^2}-\alpha \cdot \frac{c_0 \mu}{c_0(\lambda+2\mu)+\alpha^2}\right)\notag\\
		&=0.
	\end{align}
	With this estimation, we can restrict our considerations again to the plane with $y_1,y_2>0$
	\begin{align}
		&\int_{\substack{\B_\tau (0)\\y_1,y_2>0}} \frac{C_1 c_0\mu}{2\pi} \cdot \frac{y_1}{\tau^2}\left( 2-\frac{\|y\|^2}{\tau^2}\right) + \frac{\alpha C_3}{2\pi}\left( \frac{y_1(6C_4-1)}{\tau^2}-5C_4\cdot y_1 \frac{\|y\|^2}{\tau^4} \right) \ \dy \notag\\
		&=\frac{c_0 \mu C_1}{2\pi} \frac{1}{\tau^2} \int_0^{\pi/2} \int_0^\tau r \cos \ph \left(2-\frac{r^2}{\tau^2} \right) \cdot r \ \mathrm{d}r \ \mathrm{d}\ph\notag\\
		&\phantom{=}+\frac{\alpha C_3}{2\pi} \int_0^{\pi/2} \int_0^\tau \left( \frac{r \cos \ph (6C_4-1)}{\tau^2} -5C_4 r \cos \ph \cdot \frac{r^2}{\tau^4} \right) \cdot r \ \mathrm{d} r\ \mathrm{d}\ph \notag\\
		&= \frac{c_0 \mu C_1}{2\pi}\frac{1}{\tau^2} \left(\frac{2}{3}\tau^3-\frac{1}{5} \frac{\tau^5}{\tau^2}\right)+\frac{\alpha C_3}{2\pi} \left( \frac{1}{3}\frac{\tau^3}{\tau^2}(6C_4-1)-5C_4\cdot \frac{1}{5} \frac{\tau^5}{\tau^4}  \right)\notag \\
		&\to 0\ (\tau \to 0+).
	\end{align}
	Thus, we are ale to estimate the integral by a constant.
	For $\Phi_{13,\tau}^2$, it is also sufficient to have a look at the quadrant with $y_1,y_2>0$:
	\begin{align}
		\int_{\substack{\B_\tau (0)\\y_1,y_2>0}} \frac{4y_1 C_1}{\tau^3 \pi}\ \dy&=\frac{4C_1}{\tau^3 \pi}\int_0^{\pi/2} \int_0^\tau r^2 \cos \ph\ \mathrm{d} r\ \mathrm{d} \ph \notag\\
		&= \frac{4C_1}{\pi}\frac{\tau^3}{3\tau^3} \notag \\
		&= \frac{4C_1}{3 \pi}.
	\end{align}
	This integral is obviously also bounded.
	Now we have a look at the positive part of $\Phi_{31,\tau}$. The function consists of a product of a $y_1$-term and a term without $y$-dependency. Therefore, the sign of the latter term is independent of the spatial variable. Therefore we consider an arbitrary time interval and a half circle with polar coordinates together with \cref{remark:exp_integral}. We separate the spatial and the temporal integral and obtain
	\begin{align}
		\int_0^\tau \int_{t_1}^{t_2} \int_{-\pi/2}^{\pi/2} \Phi_{31,\tau} \ \mathrm{d}\ph \ \mathrm{d}t \ \mathrm{d}r
		&= \int_0^\tau \int_{t_1}^{t_2} \int_{-\pi/2}^{\pi/2} -r^2\cos \ph \ \mathrm{d}\ph
		 \exp\left( -\frac{\tau^2}{4C_2 t}\right) \\ \notag
		 &\hphantom{=}\times \left( \vphantom{\frac{\left(1-\exp \left( \frac{\tau^2}{4 C_2 t} \right)\right) }{8C_2 \mu t^2 \tau^4 \pi}}  \frac{ \alpha\mu \tau^4+8C_1(\lambda+2\mu)t\tau^2}{8C_2 \mu t^2\tau^4 \pi} + \frac{32C_1 C_2 (\lambda+2\mu)t^2 \left(1-  \exp\left( \frac{\tau^2}{4C_2 t}\right)\right) }{8C_2 \mu t^2\tau^4 \pi}\right)\ \mathrm{d}t \ \mathrm{d}r \notag \\
		&= -2\frac{\tau^3}{3}\frac{1}{8C_2 \mu \tau^4 \pi} \cdot \left[ \alpha \mu \tau^4 \left( \frac{4C_2\exp\left( -\frac{\tau^2}{4C_2 t_2}\right)}{\tau^2} -\frac{4C_2\exp\left( -\frac{\tau^2}{4C_2 t_1}\right)}{\tau^2} \right) \right. \notag\\
		& \hphantom{ -2\frac{\tau^3}{3}\frac{1}{8C_2 \mu \tau^4 \pi} \cdot  \alpha \mu \tau^4}\left.-8C_1 (\lambda+2\mu) \tau^2 \left( \Ei\left( -\frac{\tau^2}{4C_2 t_2}\right) -\Ei\left( -\frac{\tau^2}{4C_2 t_1}\right) \right) \right.\notag\\
		& \hphantom{ -2\frac{\tau^3}{3}\frac{1}{8C_2 \mu \tau^4 \pi} \cdot  \alpha \mu \tau^4}\left. +32C_1C_2 (\lambda+2\mu) \left( \frac{\tau^2}{4C_2} \left( \Ei\left( -\frac{\tau^2}{4C_2 t_2}\right)-\Ei\left( -\frac{\tau^2}{4C_2 t_1}\right)\right)\right. \right.\notag\\
		& \hphantom{ -2\frac{\tau^3}{3}\frac{1}{8C_2 \mu \tau^4 \pi} \cdot  \alpha \mu \tau^4}\left. \left. +\left(  t_2 \exp\left( -\frac{\tau^2}{4C_2 t_2}\right)-t_1\exp\left( -\frac{\tau^2}{4C_2 t_1}\right)\right)-(t_2-t_1) \right) \vphantom{\frac{4C_2\exp\left( -\frac{\tau^2}{4C_2 t_2}\right)}{\tau^2}} \right].
	\end{align}
	It is obvious that the first three lines vanish in the limit $\tau \to 0+$. Let us have a look at the last line in more detail and rearrange it to
	\begin{align}
		\underbrace{\frac{t_2}{\tau} \left(\exp\left( -\frac{\tau^2}{4C_2 t_2}\right)   -1\right)}_{\to 0 \ (\tau \to 0+)}-\underbrace{\frac{t_1}{\tau} \left(\exp\left( -\frac{\tau^2}{4C_2 t_1}\right)   -1\right)}_{\to 0\ (\tau \to 0+)}.
	\end{align}
	The usage of L'Hospital's rule gives us
	\begin{align}
		\lim_{\tau \to 0+} \frac{\exp\left( -\frac{\tau^2}{4C_2 t_1}\right)-1}{\tau}=\lim_{\tau \to 0+} \frac{\exp\left( -\frac{\tau^2}{4C_2 t_1}\right) \cdot \frac{-2\tau}{4C_2 t_1}}{1}=0.
	\end{align}
	Please note that we have a quadratic term for $t$ in the numerator of $\Phi_{31,\tau}$, that means we have a maximum of three intervals, where the function is only positive or negative.
	In the case of $\Phi_{33,\tau}$, we want to estimate the function for the space-dependent part by its maximum. Therefore, we differentiate the function with respect to $\|x\|^2$ and obtain for the numerator (which suffices for determining the zeros of the derivative)
	\begin{align}
		8t-c_0 \mu \tau^2 \begin{cases}
			>0,\ t>\frac{1}{8} c_0 \mu \tau^2,\\
			<0,\ t<\frac{1}{8}c_0 \mu \tau^2.
		\end{cases}
	\end{align}
	We start with the first case $t>\frac{1}{8}c_0 \mu \tau^2$, in which the function is monotonically increasing with respect to $\|x\|$ and attains its maximum at the boundary for $\|x\|=\tau$. We insert this maximum into the numerator of the function (here it is necessary and sufficient to have a look at the numerator to get the sign of the function). This leads us to $4t\tau^2$, which is positive and consequently also the maximum.
	In the second case it is vice versa and we have a monotonically decreasing function with its maximum in the point $(0,0)$.
	We are able to verify in the case $t<1/8 c_0 \mu \tau^2$ whether the numerator of the function is positive again:
	\begin{align}
		-4t\tau^2+c_0\mu\tau^4 > -\frac{1}{2}c_0\mu\tau^4+c_0 \mu \tau^4=\frac{1}{2} c_0 \mu \tau^4 >0.
	\end{align}
	We can continue with the estimation of the integrals for both cases. For this puspose, we split the integral over the positive part of $\Phi_{33,\tau}$ into two parts (note also \cref{remark:exp_integral}). The second part consist of the following integral, for which we estimate the maximum by $\|x\|=0$ and get (note the area of $\mathbb{B}_\tau(0)$)
	\begin{align}
		&\pi \tau^2 \int_0^{1/8 c_0 \mu \tau^2} \exp\left( -\frac{\tau^2}{4 C_2 t} \right) \frac{-4t\tau^2+c_0\mu\tau^4}{64C_2^2 t^4 \pi} \ \mathrm{d}t \notag\\
		&= \tau^2 \left[ -4\tau^2 \frac{4C_2 \exp \left(-\frac{\tau^2}{4C_2\left(\frac{1}{8} c_0 \mu \tau^2 \right)} \right) \left(\tau^2+4 \left(\frac{1}{8} c_0 \mu \tau^2 \right)C_2 \right) }{64C_2^2 \left(\frac{1}{8} c_0 \mu \tau^2 \right) \tau^4} \right.\notag\\
		&\hphantom{\tau^2  - =}\left. +c_0\mu \tau^4\frac{4C_2\exp \left(-\frac{\tau^2}{4C_2\left(\frac{1}{8} c_0 \mu \tau^2 \right)}\right) \left( 32C_2^2 \left(\frac{1}{8} c_0 \mu \tau^2 \right)^2+8C_2 \left(\frac{1}{8} c_0 \mu \tau^2 \right)\tau^2+\tau^4  \right)}{\left(\frac{1}{8} c_0 \mu \tau^2 \right)^2 \tau^6 \cdot 64C_2^2} \right].
	\end{align}
	This integral converges to a constant for $\tau \to 0+$ and is therefore bounded.
	It remains for us to discuss the integral in the first case given by $t> 1/8 \ c_0 \mu \tau^2$, for which we determined the maximum at $\|x\|=\tau$.
	\begin{align}
		\pi \tau^2 \int_{\frac{1}{8} c_0 \mu \tau^2 }^{t_0} \exp \left(-\frac{\tau^2}{4C_2 t} \right) &\frac{t\tau^2}{16 C_2^2 t^4 \pi}\ \mathrm{d}t\notag\\
		&= \frac{\tau^2}{16C_2^2} \cdot \tau^2\cdot \left[ \frac{4C_2 \exp \left(-\frac{\tau^2}{4C_2 t_0} \right)\left( \tau^2+4t_0C_2 \right)}{t_0 \tau^4}
		\right. \notag \\
		&\qquad\phantom{\frac{\tau^2}{16C_2^2} \cdot \tau^2\cdot}\left. -\frac{4C_2 \exp \left(-\frac{\tau^2}{4C_2 \left(\frac{1}{8} c_0 \mu \tau^2 \right)}\right) \left( \tau^2+4\left(\frac{1}{8} c_0 \mu \tau^2 \right)C_2 \right)}{\left(\frac{1}{8} c_0 \mu \tau^2 \right) \tau^4}
		\right],
	\end{align}
	where this integral is also bounded by a constant for $\tau \to 0+$. Here we used that $t_0$ is linked to $\tau$ by a constant $T$ in the way that $t_0=T \cdot \tau$, where $T$ is the length of our considered time interval.
\end{proof}
\noindent

\section{Conclusion}
The purpose of this article was to develop a theory for a multiscale analysis of poroelastic fields. To achieve this, we constructed scaling functions that have a physical relevance due to their relationship to the fundamental solutions of the quasi-static equations of poroelasticity. The aim was the development of a multiscale mollifier technique for the decorrelation of the relevant components displacement and pore pressure.
In the process of the mollification of the fundamental solution tensor, we did a Taylor expansion up to the linear term for the spatial component.
It turned out to be unnecessary to extend the Taylor expansion also to the time-dependence of the functions.
The application of the poroelastic differential operator to the mollified fundamental solutions led us to our desired scaling functions. From the scaling functions derived here, it is easy to construct wavelets, e.g. by inserting a sequence $\tau_j = 2^{-j}$ and building differences of scaling functions for consecutive scales $j$.
With this approach we expect to be able to decompose the data of the components at different scales to visualize underlying structures that cannot be seen in the whole picture. The foundation for the decomposition ability is given by the approximate identity. The numerical verification with real data experiments is still difficult due to the insufficient availability of data. We leave this task, therefore, to future research. 
\section{Appendix}
Here, the details of the derivation of the source scaling functions are elaborated.
We calculate the several components for the first row
\begin{align}
	\left( \Phi_{11,\tau}, \Phi_{12,\tau}, \Phi_{13,\tau} \right)
	&=L^{\pe}
	\left( u_{11,\tau}^{\CN}, u_{21,\tau}^{\CN}, p_{1,\tau}^{\St} \right)
	\notag\\
	&=
	\begin{pmatrix}
		-\frac{\lambda+\mu}{\mu} \nabla_x \left(\nabla_x \cdot \begin{pmatrix} u_{11,\tau}^{\CN} \\ u_{12,\tau}^{\CN} \end{pmatrix} \right)-\nabla_x^2 \begin{pmatrix} u_{11,\tau}^{\CN} \\ u_{12,\tau}^{\CN} \end{pmatrix}+\alpha \nabla_x p_{1,\tau}^{\St} \\
		\partial_t \left(c_0 \mu p_{1,\tau}^{\St}+\alpha\left( \nabla_x \cdot \begin{pmatrix} u_{11,\tau}^{\CN} \\ u_{12,\tau}^{\CN} \end{pmatrix} \right)\right)-\nabla_x^2 p_{1,\tau}^{\St}
	\end{pmatrix}^{\mathrm{T}}
\end{align}

and need therefore the following partial derivatives
\begin{align}
	\partial_{x_1} u_{11,\tau}^\CN&=
	\frac{C_3}{2\pi} \left[ -\frac{2x_1}{2\tau^2} +C_4 \cdot 2x_1 \cdot \left( \frac{2}{\tau^2}-\frac{\|x\|^2}{\tau^4} \right)+C_4 \cdot x_1^2 \cdot \left( -\frac{2x_1}{\tau^4}\right)\right], \\
	\partial_{x_2} u_{11,\tau}^\CN&=\frac{C_3}{2\pi} \left[ -\frac{2x_2}{2\tau^2} +C_4 \cdot x_1^2 \cdot \left( -\frac{2x_2}{\tau^4}\right)\right], \\
	\partial_{x_1^2} u_{11,\tau}^\CN&= \frac{C_3}{2\pi} \left( -\frac{1}{\tau^2}+C_4 \cdot \left( \frac{4}{\tau^2}-\frac{12x_1^2+2x_2^2}{\tau^4} \right) \right), \\
	\partial_{x_2^2}u_{11,\tau}^\CN&=\frac{C_3}{2\pi} \left[ -\frac{1}{\tau^2} +C_4 \cdot x_1^2 \cdot \left( -\frac{2}{\tau^4}\right)\right], \\
	\partial_{x_1} \partial_{x_2} u_{11,\tau}^\CN &= \frac{C_3 C_4}{2\pi} \left( -\frac{4x_1x_2}{\tau^4} \right),  \\
	\partial_{x_1} u_{12,\tau}^\CN&=\frac{C_3 C_4}{2\pi} \left[  x_2 \cdot \left( \frac{2}{\tau^2}-\frac{\|x\|^2}{\tau^4} \right) +x_1x_2 \cdot \left(-\frac{2x_1}{\tau^4}\right) \right],\\
	\partial_{x_2} u_{12,\tau}^\CN&=\frac{C_3 C_4}{2\pi} \left[  x_1 \cdot \left( \frac{2}{\tau^2}-\frac{\|x\|^2}{\tau^4} \right) +x_1x_2 \cdot \left(-\frac{2x_2}{\tau^4}\right) \right],\\	
	\partial_{x_1^2} u_{12,\tau}^\CN &= \frac{C_3 C_4}{2\pi} \left( -\frac{6x_1x_2}{\tau^4} \right),\\
	\partial_{x_2^2} u_{12,\tau}^\CN &= \frac{C_3 C_4}{2\pi} \left( -\frac{6x_1x_2}{\tau^4} \right),\\	
	\partial_{x_1} \partial_{x_2} u_{12,\tau}^\CN&= \frac{C_3 C_4}{2\pi} \left( \frac{2}{\tau^2}-\frac{3 \|x\|^2}{\tau^4} \right).
\end{align}
Now we can assemble the several components of our source scaling functions and obtain

\begin{align}
	\renewcommand{\arraystretch}{1.3}	
	\nabla_x p_{1,\tau}^{\St}&=\frac{C_1}{2\pi} \frac{1}{\tau^2} \begin{pmatrix}
		2-\frac{3x_1^2+x_2^2}{\tau^2} \\ -\frac{2x_1x_2}{\tau^2}
	\end{pmatrix},
	\renewcommand{\arraystretch}{1.0} \\
	\nabla_x^2 u_{11,\tau}^{\CN}&=\frac{C_3}{2\pi}\left( -\frac{2}{\tau^2}+C_4 \cdot \left(- \frac{12x_1^2+2x_2^2+2x_1^2}{\tau^4}+\frac{4}{\tau^2} \right) \right) \notag\\
	&=\frac{C_3}{2\pi}\left( \frac{-2+4C_4}{\tau^2}+C_4 \cdot \left(- \frac{14x_1^2+2x_2^2}{\tau^4} \right) \right), \\
	\nabla_x^2 u_{12,\tau}^{\CN}&=-\frac{C_3 C_4}{\pi} \frac{6x_1x_2}{\tau^4},\\
	\nabla_x \left(\nabla_x\cdot \begin{pmatrix}
		u_{11,\tau}^{\CN} \\ u_{12,\tau}^{\CN}
	\end{pmatrix} \right)&=\begin{pmatrix}
		\partial_{x_1^2} u_{11,\tau}^{\CN}+\partial_{x_1} \partial_{x_2} u_{12,\tau}^{\CN} \\
		\partial_{x_2}\partial_{x_1} u_{11,\tau}^{\CN}+\partial_{x_2^2} u_{12,\tau}^{\CN}
	\end{pmatrix} \notag \\
	&= \renewcommand{\arraystretch}{1.3}	\begin{pmatrix}
		\frac{C_3}{2\pi} \left( -\frac{1}{\tau^2}+C_4 \cdot \left( \frac{6}{\tau^2}-\frac{15x_1^2+5x_2^2}{\tau^4} \right) \right)\\
		\frac{C_3 C_4}{2\pi} \left( -\frac{10x_1x_2}{\tau^4} \right)
	\end{pmatrix}.
	\renewcommand{\arraystretch}{1.0}
\end{align}

A further summary of the components above leads us to
\begin{align}
	\Phi_{11,\tau}(x)&= -\frac{\lambda+\mu}{\mu} \frac{C_3}{2\pi} \left( -\frac{1}{\tau^2}+C_4 \cdot \left( \frac{6}{\tau^2}-\frac{15x_1^2+5x_2^2}{\tau^4} \right) \right)\notag\\
	&\phantom{=}\ -\frac{C_3}{2\pi} \left( \frac{-2+4C_4}{\tau^2}+C_4 \cdot \left( -\frac{14x_1^2+2x_2^2}{\tau^4} \right)  \right)+\alpha \frac{C_1}{2\pi} \frac{1}{\tau^2} \left( 2-\frac{3x_1^2+x_2^2}{\tau^2} \right), \\
	\Phi_{12,\tau}(x)&=-\frac{\lambda+\mu}{\mu} \cdot \frac{C_3 C_4}{2\pi} \left(-\frac{10 x_1x_2}{\tau^4}\right)+ \frac{C_3 C_4}{\pi} \cdot \frac{6x_1 x_2}{\tau^4}-\alpha \frac{C_1}{2\pi} \frac{2 x_1 x_2}{\tau^4} \notag\\
	&= \frac{x_1x_2}{\tau^4 \pi} \left( \frac{\lambda+\mu}{\mu} \cdot 5 C_3 C_4  +6C_3 C_4-\alpha C_1 \right).
\end{align}
For $\Phi_{13,\tau}$, we have to consider that the whole entries in the fundamental solution tensor belonging to $\boldsymbol{u}^\CN$ and $p^\St$ are equipped with $\delta_t$. Furthermore we have to observe that the third equation contains a derivative with respect to $t$. The Laplacian of $p^\St_{1,\tau}$ is given by
\begin{align}
	\nabla_x^2 p_{1,\tau}^\St &= -\frac{4x_1 C_1}{\tau^4 \pi}
\end{align}
and together with the preliminary considerations from above, we get for the two components of $\Phi_{13,\tau}$
\begin{align}
	\Phi_{13,\tau}^1(x)&=\left[ \frac{c_0 \mu C_1}{2\pi}\cdot \frac{x_1}{\tau^2}\left( 2-\frac{\|x\|^2}{\tau^2}\right)+\frac{\alpha C_3}{2\pi} \left( \frac{x_1 (6C_4-1)}{\tau^2}-5C_4 \cdot x_1\cdot \frac{\|x\|^2}{\tau^4} \right) \right],\\
	\Phi_{13,\tau}^2(x)&=\frac{4 x_1 C_1}{\tau^4 \pi}.
\end{align}
The first component is the part equipped with $\delta_t'$ in the whole entry of the source scaling function tensor and the second one is the factor of $\delta_t$.
Here it is necessary to do a little modification of $\Phi_{13,\tau}^2$ to guarantee that the property of an approximate identity holds true. For this purpose, we will change the $\tau^4$ in the denominator to $\tau^3$, which can be seen as a part of the regularization.
We continue with the second row of $\Phi_{\tau}$, where we have that the calculations are similar to those above, because of the symmetry. We get
\begin{align}
	\begin{pmatrix}
		\Phi_{21,\tau},  \Phi_{22,\tau}, \Phi_{23,\tau}
	\end{pmatrix}
	&=L^{\pe} \begin{pmatrix}
		u_{21,\tau}^{\CN}, u_{22,\tau}^{\CN}, p_{2,\tau}^{\St}
	\end{pmatrix}  \notag \\
	&=
	\begin{pmatrix}
		-\frac{\lambda+\mu}{\mu} \nabla_x \left(\nabla_x \cdot \begin{pmatrix} u_{21,\tau}^{\CN} \\ u_{22,\tau}^{\CN} \end{pmatrix} \right)-\nabla_x^2 \begin{pmatrix} u_{21,\tau}^{\CN} \\ u_{22,\tau}^{\CN} \end{pmatrix}+\alpha \nabla_x p_{2,\tau}^{\St} \\
		\partial_t \left(c_0 \mu p_{2,\tau}^{\St}+\alpha\left( \nabla_x \cdot \begin{pmatrix} u_{21,\tau}^{\CN} \\ u_{22,\tau}^{\CN} \end{pmatrix} \right)\right)-\nabla_x^2 p_{2,\tau}^{\St}
	\end{pmatrix}^{\mathrm{T}}.
\end{align}

In analogy to above, he have for the derivatives for $\|x\|<\tau$
\begin{align}
	\nabla_x p_{2,\tau}^{\St}&=\frac{C_1}{2\pi} \frac{1}{\tau^2} \begin{pmatrix}
		-\frac{2x_1x_2}{\tau^2} \\ 2-\frac{3x_2^2+x_1^2}{\tau^2}
	\end{pmatrix}, \\
	\nabla_x^2 u_{21,\tau}^{\CN} &=-\frac{C_3 C_4}{\pi} \cdot \frac{6x_1 x_2}{\tau^4}, \\
	\nabla_x^2 u_{22,\tau}^{\CN}&=\frac{C_3}{2\pi}\left( -\frac{2}{\tau^2}+C_4 \cdot \left( -\frac{14x_2^2+2x_1^2-4}{\tau^4} \right) \right), \\
	\nabla_x\left(\nabla_x\cdot \begin{pmatrix}
		u_{21,\tau}^{\CN} \\ u_{22,\tau}^{\CN}
	\end{pmatrix} \right)
	&= \renewcommand{\arraystretch}{1.3}	\begin{pmatrix}
		\frac{C_3 C_4}{2\pi} \left( -\frac{10x_1x_2}{\tau^4} \right)\\
		\frac{C_3}{2\pi} \left( -\frac{1}{\tau^2}+C_4 \cdot \left( \frac{6}{\tau^2}-\frac{5x_1^2+15x_2^2}{\tau^4} \right) \right)
	\end{pmatrix}.
	\renewcommand{\arraystretch}{1.0}
\end{align}
Since we have symmetric relations between the functions in the first and second row (changed roles of $x_1$ and $x_2$ ), we do not write them down here separately. In the case of $\Phi_{23,\tau}^2$, we do the same modification in the denominator as we did it for $\Phi_{13,\tau}^2$.
Let us continue with the third row, that means the space and time dependent functions.
\begin{align}
	\begin{pmatrix}
		\Phi_{31,\tau}, \Phi_{32,\tau}, \Phi_{33,\tau}
	\end{pmatrix}
	&=L^{\pe} \begin{pmatrix}
		u_{1,\tau}^{\Si}, u_{2,\tau}^{\Si}, p_{\tau}^{\Si}
	\end{pmatrix} \notag \\
	&=
	\begin{pmatrix}
		-\frac{\lambda+\mu}{\mu} \nabla_x \left(\nabla_x \cdot \begin{pmatrix} u_{1,\tau}^{\Si} \\ u_{2,\tau}^{\Si} \end{pmatrix} \right)-\nabla_x^2 \begin{pmatrix} u_{1,\tau}^{\Si} \\ u_{2,\tau}^{\Si} \end{pmatrix}+\alpha \nabla_x p_{\tau}^{\Si} \\
		\partial_t \left(c_0 \mu p_{\tau}^{\Si}+\alpha\left( \nabla_x \cdot \begin{pmatrix} u_{1,\tau}^{\Si} \\ u_{2,\tau}^{\Si} \end{pmatrix} \right)\right)-\nabla_x^2 p_{\tau}^{\Si}
	\end{pmatrix}^{\mathrm{T}}.
\end{align}

The calculation involves the following derivatives:
\begin{align}
	\nabla_x \left(\nabla_x \cdot \begin{pmatrix}
		x_1 \|x\|^2 \\
		x_2 \|x\|^2
	\end{pmatrix} \right)
	&=\nabla_x \left(\nabla_x \cdot
	\begin{pmatrix}
		x_1 (x_1^2+x_2^2) \\
		x_2 (x_1^2+x_2^2)
	\end{pmatrix} \right)
	=\begin{pmatrix}
		8x_1\\
		8x_2
	\end{pmatrix},\\
	\nabla_x\left(\nabla_x \cdot \begin{pmatrix}
		u_{1,\tau}^{\Si}\\
		u_{2,\tau}^{\Si}
	\end{pmatrix} \right)
	&= \frac{8C_1}{2\pi} \cdot8 \cdot \begin{pmatrix}
		x_1 \\ x_2 \end{pmatrix} \left( -\frac{1}{\tau^4}+ \exp \left( -\frac{\tau^2}{4 C_2 t} \right)\frac{1}{4 C_2 t}\frac{1}{\tau^2} \right. \notag \\
	&\phantom{ =\frac{8C_1}{2\pi} \cdot8 \cdot \begin{pmatrix}
			x_1 \\ x_2 \end{pmatrix} \bigg( } \left. +\frac{1}{\tau^4} \exp \left( -\frac{\tau^2}{4 C_2 t} \right)\right), \\
	\nabla_x^2
	\begin{pmatrix}
		x_1 \|x\|^2 \\
		x_2 \|x\|^2
	\end{pmatrix}
	&=\nabla_x^2  \begin{pmatrix}
		x_1 (x_1^2+x_2^2) \\
		x_2 (x_1^2+x_2^2)
	\end{pmatrix} = \begin{pmatrix}
		8x_1\\
		8x_2
	\end{pmatrix},\\
	\nabla_x^2 u_{\tau}^{\Si}&= \frac{C_1}{2\pi} \cdot8 \cdot \begin{pmatrix}
		x_1 \\ x_2 \end{pmatrix} \left( -\frac{1}{\tau^4}+ \exp \left( -\frac{\tau^2}{4 C_2 t} \right)\frac{1}{4 C_2 t}\frac{1}{\tau^2} \right. \notag \\
	&\phantom{ =\frac{C_1}{2\pi} \cdot8 \cdot \begin{pmatrix}
			x_1 \\ x_2 \end{pmatrix} \bigg( } \left.+\frac{1}{\tau^4} \exp \left( -\frac{\tau^2}{4 C_2 t} \right)\right),\\
	\nabla_x p_\tau^{\Si}&= \frac{1}{4\pi t}  \exp \left( -\frac{\tau^2}{4 C_2 t} \right) \cdot \left( -\frac{1}{4C_2 t}\right) \cdot \begin{pmatrix}
		2x_1 \\2x_2
	\end{pmatrix}\notag\\
	&= -\begin{pmatrix}
		x_1 \\ x_2
	\end{pmatrix}
	\exp \left( -\frac{\tau^2}{4 C_2 t} \right) \frac{1}{8 \pi C_2 t^2}. \label{eq:grad_p1St}
\end{align}
Here the $x \|x\|$-term is the relevant term that remains if we consider second-order derivatives.
Putting together these derivatives, we get the first two entries $\Phi_{31,\tau}$ and $\Phi_{32,\tau}$, which are
\begin{align}
	\begin{pmatrix}
		\Phi_{31,\tau} \\ \Phi_{32,\tau}
	\end{pmatrix}&=\left[-\frac{\lambda+\mu}{\mu}-1\right] \frac{C_1}{2\pi} \cdot8 \cdot \begin{pmatrix}
		x_1 \\ x_2
	\end{pmatrix} \left( -\frac{1}{\tau^4}+ \exp \left( -\frac{\tau^2}{4 C_2 t} \right)\frac{1}{4 C_2 t}\frac{1}{\tau^2}\right. \notag \\
	&\phantom{=}\left.  +\frac{1}{\tau^4} \exp \left( -\frac{\tau^2}{4 C_2 t} \right)\right)
	-\begin{pmatrix}
		x_1 \\ x_2
	\end{pmatrix}\frac{\alpha }{8C_2 \pi t^2} \exp \left( -\frac{\tau^2}{4 C_2 t} \right) \notag\\
	&=  \begin{pmatrix}
		x_1 \\ x_2
	\end{pmatrix} \frac{(-\lambda-2\mu)8C_1  \left(-4C_2t^2+t\tau^2\exp \left( -\frac{\tau^2}{4 C_2 t} \right)+4C_2t^2 \exp \left( -\frac{\tau^2}{4 C_2 t} \right) \right)}{8C_2 \mu t^2 \tau^4 \pi}\notag\\
	&\phantom{=}- \begin{pmatrix}
		x_1 \\ x_2
	\end{pmatrix} \frac{\alpha  \mu \tau^4\exp \left(-\frac{\tau^2}{4 C_2 t} \right)}{8C_2 \mu t^2 \tau^4 \pi}\notag\\
	&= -\begin{pmatrix}
		x_1 \\ x_2
	\end{pmatrix} \exp \left( -\frac{\tau^2}{4 C_2 t} \right)  \left( \vphantom{\frac{\left(1-\exp \left( \frac{\tau^2}{4 C_2 t} \right)\right) }{8C_2 \mu t^2 \tau^4 \pi}} \frac{\alpha\mu \tau^4+8C_1(\lambda+2\mu)t\tau^2}{8C_2 \mu t^2 \tau^4 \pi} \right. \notag  \\
	&\qquad \hphantom{ -\begin{pmatrix}
			x_1 \\ x_2
		\end{pmatrix} \exp \left( -\frac{\tau^2}{4 C_2 t} \right)}\left.+ \frac{32C_1C_2(\lambda+2\mu)t^2\left(1-\exp \left( \frac{\tau^2}{4 C_2 t} \right)  \right)}{8C_2 \mu t^2 \tau^4 \pi}   \right) .
\end{align}
Now the function $\Phi_{33,\tau}$ is left. The calculation of the Laplacian of $p^\Si_{1,\tau}$ requires the use of the gradient from above and the application of the divergence. The several components are the following
\begin{align}
	\nabla_x^2 p^{\Si}_\tau &= \mathrm{div} \left( -\begin{pmatrix}
		x_1 \\ x_2
	\end{pmatrix}
	\exp \left( -\frac{\tau^2}{4 C_2 t} \right) \frac{1}{8 \pi C_2 t^2}\right) \notag\\
	&= \frac{-1}{4 \pi C_2 t^2} \exp \left( -\frac{\tau^2}{4 C_2 t} \right), \\
	\partial_t p^{\Si}_\tau &=\frac{-1}{4 \pi t^2} \exp \left( -\frac{\tau^2}{4 C_2 t} \right) \left[ 1-\frac{1}{4C_2 t} (\|x\|^2-\tau^2) \right]\notag\\
	&\phantom{=}+\frac{1}{4\pi t } \exp \left( -\frac{\tau^2}{4 C_2 t} \right) \cdot \frac{\tau^2}{4 C_2 t^2} \left[ 1-\frac{1}{4C_2 t} (\|x\|^2-\tau^2) \right]\notag\\
	&\phantom{=}+\frac{1}{4 \pi t}\exp \left( -\frac{\tau^2}{4 C_2 t} \right)\cdot \frac{1}{4 C_2 t^2} (\|x\|^2-\tau^2)\notag\\
	&=\exp \left( -\frac{\tau^2}{4 C_2 t} \right) \left[-\frac{1}{4\pi t^2}+\frac{1}{16C_2 t^3\pi} (\|x\|^2-\tau^2)+\frac{\tau^2}{16 C_2 t^3 \pi}\right.\notag \\
	&\hphantom{\exp \left( -\frac{\tau^2}{4 C_2 t} \right)=-}\left.-\frac{\tau^2}{64C_2^2 t^4 \pi} (\|x\|^2-\tau^2)+\frac{1}{16C_2 t^3 \pi} (\|x\|^2-\tau^2)  \right] \notag \\
	&= \exp \left( -\frac{\tau^2}{4 C_2 t} \right) \left[ -\frac{1}{4\pi t^2} +\frac{1}{16C_2t^3 \pi} (2\|x\|^2-\tau^2)-\frac{\tau^2}{64 C_2^2 t^4 \pi} (\|x\|^2-\tau^2)  \right],\\
	\partial_t (\nabla_x \cdot u^{\Si})&= \partial_t \left( \frac{C_1}{2\pi} \left[ 2\cdot \left[\frac{1}{\tau^2}-\frac{1}{\tau^2} \exp \left( -\frac{\tau^2}{4 C_2 t} \right) \right.  \right. \right.\notag\\
	&\phantom{\partial_t=-}\left .\left.  \left.+ \left( \frac{-1}{ \tau^4}+\frac{1}{4 C_2 t} \exp \left( -\frac{\tau^2}{4 C_2 t} \right)\cdot \frac{1}{\tau^2}+\frac{1}{ \tau^4} \exp \left(-\frac{\tau^2}{4 C_2 t}  \right) \right)(\|x\|^2-\tau^2) \right]  \right. \right.  \notag\\
	&\phantom{\partial_t=-}\left. \left. +2\|x\|^2 \left[ \frac{-1}{\tau^4}+\frac{1}{4C_2 t} \frac{1}{\tau^2}  \exp \left( -\frac{\tau^2}{4 C_2 t} \right)+\frac{1}{\tau^4}  \exp \left( -\frac{\tau^2}{4 C_2 t} \right) \right] \right] \right)\notag\\
	&= \frac{C_1}{ \pi} \left[-\frac{1}{\tau^2} \exp \left( -\frac{\tau^2}{4 C_2 t} \right) \cdot\frac{\tau^2}{4C_2 t^2}+\left(-\frac{1}{4C_2 t^2} \exp \left( -\frac{\tau^2}{4 C_2 t} \right) \cdot \frac{1}{\tau^2} \right. \right.\notag\\
	&\phantom{\frac{C_1}{ \pi} -=}\left.\left.  +\frac{1}{4C_2t}  \exp \left( -\frac{\tau^2}{4 C_2 t} \right) \cdot \frac{\tau^2}{4C_2 t^2}\cdot \frac{1}{\tau^2}+\frac{1}{\tau^4}  \exp \left( -\frac{\tau^2}{4 C_2 t} \right) \cdot \frac{\tau^2}{4 C_2t^2} \right) \right. \notag\\
	&\phantom{\frac{C_1}{ \pi}- =}\left. \times ( 2\|x\|^2-\tau^2)\vphantom{\frac{1}{\tau^2}} \right] \notag\\
	&= \frac{C_1}{ \pi} \exp \left( -\frac{\tau^2}{4 C_2 t} \right)\left[-\frac{1}{4C_2t^2}   +\frac{1}{16C_2^2t^3} ( 2\|x\|^2-\tau^2)\right] .
\end{align}

Putting together the derivatives and inserting the material constants given in \eqref{eq:C1C2}, we obtain
\begin{align}
	\Phi_{33,\tau}&=c_0 \mu \exp \left( -\frac{\tau^2}{4 C_2 t} \right) \left[ -\frac{1}{4\pi t^2}+\frac{1}{16C_2 t^3 \pi}(2\|x\|^2-\tau^2)-\frac{\tau^2}{64C_2^2t^4 \pi} (\|x\|^2-\tau^2) \right] \notag\\
	&\phantom{=}+ \frac{\alpha C_1}{\pi} \exp \left( -\frac{\tau^2}{4 C_2 t} \right)\left[ -\frac{1}{4C_2t^2} +\frac{1}{16C_2^2t^3} ( 2\|x\|^2-\tau^2)\right] \notag \\
	&\phantom{=}+\exp \left( -\frac{\tau^2}{4 C_2 t} \right) \frac{1}{4C_2t^2 \pi}\notag\\
	&= \exp \left( -\frac{\tau^2}{4 C_2 t} \right) \left[ \frac{\overbrace{-C_2c_0\mu-C_1\alpha}^{=-1}}{4C_2 \pi t^2}+\frac{\overbrace{C_2c_0\mu+C_1\alpha}^{=1}}{16C_2^2t^3 \pi}(2\|x\|^2-\tau^2) \right. \notag \\
	&\quad \phantom{=\exp \left( -\frac{\tau^2}{4 C_2 t} \right)}\left. -\frac{c_0 \mu \tau^2}{64C_2^2t^4 \pi}(\|x\|^2-\tau^2)+\frac{1}{4C_2t^2 \pi} \vphantom{\frac{\overbrace{-C_2c_0\mu-C_1\alpha}^{=-1}}{4C_2 \pi t^2}} \right]\notag\\
	&= \exp \left( -\frac{\tau^2}{4 C_2 t} \right) \frac{4t(2\|x\|^2-\tau^2)+c_0\mu \tau^2(\tau^2-\|x\|^2)}{64C_2^2 t^4 \pi}.
\end{align} 
\bibliographystyle{abbrv}
\bibliography{biblio}
\end{document}